\documentclass[11pt]{article}

\pdfoutput=1

\newcommand{\W}{\mathcal{W}}

\newcommand*{\tran}{^{\mkern-1.5mu\mathsf{T}}}

\def \grad{\nabla}
\def \p{\partial}

\def \CC{\mathbb{C}}

\def \FF{\mathbb{F}}

\def \II{\mathbb{I}}

\def \PP{\mathbb{P}}

\def \FFA{\mathbb{F}_\text{A}}
\def \FFE{\mathbb{F}_\text{E}}

\newcommand{\gf}[1]{\gamma_\text{#1}}
\newcommand{\fiber}[2]{\mathbf{#1}_\text{#2}}
\newcommand{\dyad}[4]{\fiber{#1}{#2}\otimes\fiber{#3}{#4}}

\def \F{\mathbf{F}}

\def \N{\mathbf{N}}

\def \U{\mathbf{U}}
\def \X{\mathbf{X}}
\def \f{\mathbf{f}}

\def \n{\mathbf{n}}

\def \vChi{\boldsymbol{\chi}}
\def \vchi{\boldsymbol{\chi}}

\def \u{\mathbf{u}}
\def \x{\mathbf{x}}

\def \Omegas{\Omega^\text{s}}
\def \Omegaf{\Omega^\text{f}}

\def \Dx{{\mathrm d} \x}

\def \DX{{\mathrm d} \X}

\usepackage{booktabs}
\usepackage[margin=15pt,font=small,labelfont={bf,sf},justification=justified]{caption}
\usepackage[superscript,biblabel]{cite}
\usepackage{color}
\usepackage{float}
\usepackage{graphicx}
\usepackage[bbgreekl]{mathbbol}
\usepackage{mathtools}
\usepackage{multibib}
\usepackage{multirow}
\usepackage{rotating}
\usepackage{tabularx}
\usepackage{titleref}
\usepackage{url}

\usepackage[scaled=.98,p]{XCharter}
\usepackage[scaled=1.04,varqu,varl]{inconsolata}
\usepackage[type1]{cabin}
\usepackage[uprightscript,xcharter,vvarbb,scaled=1.05]{newtxmath}
\usepackage[cal=euler,scr=boondoxo,bb=boondox]{mathalpha}
\usepackage[T1]{fontenc}
\usepackage[final,protrusion=true,expansion=true]{microtype}

\usepackage{titling}
\usepackage{authblk} % note: titling needs to be loaded before authblk
\pretitle{\begin{center}\bfseries\sffamily\LARGE}
\posttitle{\end{center}}

\predate{\begin{center}\normalsize}
\postdate{\end{center}}
\usepackage{abstract}
\usepackage{sectsty} % note: the main alternative package, titlesec, does not work with titleref
\allsectionsfont{\sffamily}

\usepackage[left=0.5in,right=0.5in,top=0.5in,bottom=0.5in,includefoot,heightrounded]{geometry}

\allowdisplaybreaks

\newcites{supp}{METHODS REFERENCES}

\newcommand{\beginsupplement}{%
        \setcounter{table}{0}
        \renewcommand{\thetable}{S\arabic{table}}%
        \setcounter{figure}{0}
        \renewcommand{\thefigure}{S\arabic{figure}}%
        \setcounter{section}{0}
        \renewcommand{\thesection}{S\arabic{section}}
        \setcounter{equation}{0}
        \renewcommand{\theequation}{S\arabic{equation}}
}

\title{Simulating Cardiac Fluid Dynamics in the Human Heart}
\author[1,*]{Marshall~Davey}
\author[2,*]{Charles~Puelz}
\author[3]{Simone~Rossi}
\author[3]{Margaret~Anne~Smith}
\author[3]{David~R.~Wells}
\author[4]{Gregory~M.~Sturgeon}
\author[4]{W.~Paul~Segars}
\author[5]{John~P.~Vavalle}
\author[6]{Charles~S.~Peskin}
\author[7--10]{Boyce~E.~Griffith}
\affil[1]{Curriculum in Bioinformatics and Computational Biology, University of North Carolina, Chapel Hill, NC, USA}
\affil[2]{Department of Pediatrics-Cardiology, Baylor College of Medicine and Texas Children's Hospital, Houston, TX, USA}
\affil[3]{Department of Mathematics, University North Carolina, Chapel Hill, NC, USA}
\affil[4]{Department of Radiology, Duke University Medical Center, Durham, NC, USA}
\affil[5]{Division of Cardiology, Department of Medicine, University of North Carolina School of Medicine, Chapel Hill, NC, USA}
\affil[6]{Courant Institute of Mathematical Sciences, New York University, New York, NY, USA}
\affil[7]{Departments of Mathematics and Biomedical Engineering, University of North Carolina, Chapel Hill, NC, USA}
\affil[8]{Carolina Center for Interdisciplinary Applied Mathematics, University of North Carolina, Chapel Hill, NC, USA}
\affil[9]{Computational Medicine Program, University of North Carolina School of Medicine, Chapel Hill, NC, USA}
\affil[10]{McAllister Heart Institute, University of North Carolina School of Medicine, Chapel Hill, NC, USA\vspace{\baselineskip}}
\affil[*]{These authors contributed equally to this work}

\begin{document}
\maketitle

\begin{abstract}
\noindent Cardiac fluid dynamics fundamentally involves interactions between complex blood flows and the structural deformations of the muscular heart walls and the thin, flexible valve leaflets.
There has been longstanding scientific, engineering, and medical interest in creating mathematical models of the heart that capture, explain, and predict these fluid-structure interactions.
However, existing computational models that account for interactions among the blood, the actively contracting myocardium, and the cardiac valves are limited in their abilities to predict valve performance, resolve fine-scale flow features, or use realistic descriptions of tissue biomechanics.
Here we introduce and benchmark a comprehensive mathematical model of cardiac fluid dynamics in the human heart.
A unique feature of our model is that it incorporates biomechanically detailed descriptions of all major cardiac structures that are calibrated using tensile tests of human tissue specimens to reflect the heart's microstructure.
Further, it is the first fluid-structure interaction model of the heart that provides anatomically and physiologically detailed representations of all four cardiac valves.
We demonstrate that this integrative model generates physiologic dynamics, including realistic pressure-volume loops that automatically capture isovolumetric contraction and relaxation, and predicts fine-scale flow features.
None of these outputs are prescribed; instead, they emerge from interactions within our comprehensive description of cardiac physiology.
Such models can serve as tools for predicting the impacts of medical devices or clinical interventions.
They also can serve as platforms for mechanistic studies of cardiac pathophysiology and dysfunction, including congenital defects, cardiomyopathies, and heart failure, that are difficult or impossible to perform in patients.
\end{abstract}


\section*{Introduction}

The heart is the most dynamic organ in the body and has been a focus of scientific and medical inquiry for millennia.
Studies of the human heart began with detailed descriptions of its gross anatomy and have evolved to encompass a diverse set of research questions and approaches \cite{Clayton12}.
Current studies vary widely in both methodology and scale, from wet-lab experiments of cell function to analytic characterization of muscle fiber orientation \cite{Radisic04, Savadjiev12}.
Animal models were some of the first systems used to understand the heart as a dynamic system in vivo \cite{Ericsson13, Camacho16}, but they are limited by the invasive nature of the experimental measurements, which impact the observed dynamics, as well as by the differences in anatomy and physiology of other animals as compared to humans.
Imaging and catherization studies provide a means to study human cardiac function in vivo, but although technological advancements continue to improve these approaches, they can be costly, are limited in the detail of their measurements, and can pose risks to human subjects.
Further, whereas methods for measuring cardiac function can assess the present state of the heart, models are critical for predicting future states of the heart following growth, remodeling, or clinical intervention.

Predictive mathematical models of the heart can capture important features of cardiac function and serve as platforms for treatment planning and medical device design \cite{Morrison23}.
Despite substantial scientific, engineering, and medical research, however, previous computer models of the heart that account for interactions among the blood, the actively contracting myocardium, and the cardiac valves have included important simplifications that impact their ability to predict valve performance, resolve fine-scale flow features, or use realistic models of tissue biomechanics.
To our knowledge, the earliest four-chambered heart model was developed by Peskin and McQueen \cite{Peskin96, McQueen00} using the immersed boundary method \cite{Peskin02}.
Although their model captured the complex interactions of the heart muscle, valves, and blood, its anatomy was highly idealized, and it described the biomechanics of the heart using systems of one-dimensional elastic fibers that are challenging to calibrate to human data.
Baillargeon et al.~\cite{Baillargeon14} constructed a model that included detailed descriptions of the myocardium coupled to an electrophysiology model for the four heart chambers.
However, to obtain boundary conditions for the chamber walls, they used a simplified fluid model that neglected spatial variations in the pressure and velocity fields.
Fedele et al.~\cite{Fedele22} created a biomechanically detailed four-chambered electromechanical heart model coupled to a zero-dimensional blood circulation model via pressure boundary conditions.
Their model produced pressure-volume loops in the atria and ventricles in good agreement with clinical data but also did not capture fine-scale flow features in the chambers and around valve leaflets because of its simplified treatment of the intracardiac fluid mechanics.
Kariya et al.~\cite{Kariya20} built a four-chambered heart model based on the arbitrary Lagrangian-Eulerian method.
Their construction included models for oxygen transport, electrophysiology, and the valve leaflets. However, their approach used a simplified biomechanics model that neglects nonlinear and anisotropic responses of real heart valves and required the explicit modeling of contact between structures.

Here we introduce and benchmark a new comprehensive fluid-structure interaction (FSI) model of the human heart.
The model anatomy is derived from cardiac computed tomography (CT) imaging and includes fully three-dimensional descriptions of all major cardiac structures, including the atria, ventricles, mitral and tricuspid valves and their chordae tendineae and papillary muscles, aortic and pulmonary valves, and great vessels.
The biomechanical models of the heart and its valves are parameterized using experimental tensile test data obtained exclusively from human tissue specimens \cite{Augustin19, Gultekin16, Lim80, Pham17, Zuo16}.
Model-based approaches \cite{Hasan17, Rossi14, Rossi22} consistent with earlier experimental work \cite{Arts01, Bigi82, Driessen05, Ho12, Nielsen91, Streeter73} describe the heart's fiber architecture.
FSI simulations use the immersed finite element/finite difference (IFED) version \cite{Griffith17, Lee22, VadalaRoth20, Wells23} of the immersed boundary method \cite{Peskin02}, which automatically handles contact between structures, including the valve leaflets \cite{Brown23, Hasan17, Lee20, Lee21}.
Several recent methodological developments enabled this model, including modern tetrahedral mesh generation techniques \cite{Hu20} and stabilized nodal IFED methods \cite{VadalaRoth20, Wells23}.
Our model generates physiological stroke volumes, pressure-volume loops, and valvular pressure-flow relationships, illustrating its potential for predicting cardiac function in both health and disease.

\section*{Results}

\subsection*{Modeling Human Cardiac Anatomy and Physiology}

\begin{figure}[t!]
\begin{center}
\includegraphics[width=0.9\textwidth]{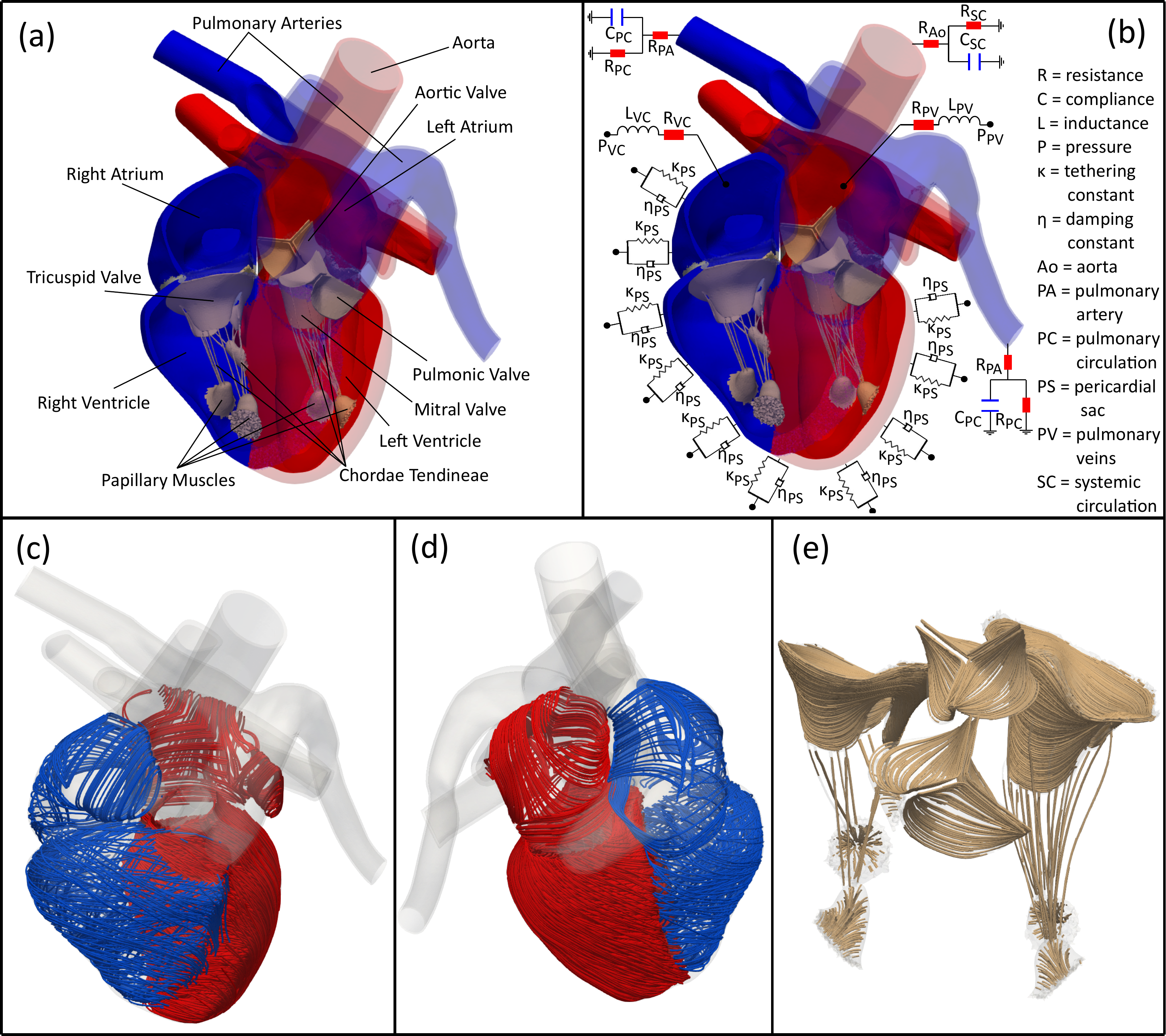}
\end{center}
\captionsetup{width=0.9\textwidth}
\caption{Anatomical and physiological aspects of the heart model.
Panel (a) visualizes the structural components of the heart model.
The anterior portions of the chambers and great vessels are transparent to reveal the four valves and valvular complexes.
Panel (b) provides a schematic of the reduced order models used for the pericardial sac, the systemic and pulmonary circulations, and the venous return to the atria.
Panels (c) and (d) depict the main myocardial fibers from two different views with the right heart and left heart chambers depicted in blue and red, respectively, and panel (e) visualizes the main fiber directions on the valve leaflets, chordae, and papillary muscles.
The fiber coloration was chosen for visual clarity.}
\label{fig:model-construction}
\end{figure}

The defining feature of the present model is that all of its dynamics emerge from interactions among its components.
We prescribe only the anatomy and physiology, including tissue properties, muscle activation timings, and physiological boundary conditions.
Fig.~\ref{fig:model-construction} provides an overview of the model and Fig.~\ref{fig:meshViz} shows the finite element mesh used in our simulations.

The anatomy of the heart chambers and the nearby great vessels were reconstructed from deidentified cardiac CT images of a healthy adult male provided by Siemens Healthineers.
The images used to reconstruct the model correspond to the early diastolic phase of the cardiac cycle, when the heart is in its most relaxed state \cite{Palit17}.
It can be difficult or impossible to capture the valve leaflets and chordae tendineae from whole-heart CT images \cite{DiCarli16}, and, indeed, the images used in this construction do not clearly show these structures.
Consequently, we generated idealized anatomical models of the valve leaflets based on dimensions obtained from studies of human valves and merged these with the CT-derived chamber anatomy.

Cardiac tissues are highly anisotropic.
In the myocardium, this a consequence of the alignment and organization of the muscle fibers \cite{Holzapfel09}.
Likewise, the mechanical behavior of the valve leaflets is characterized by families of aligned collagen fibers \cite{Pham17}.
To capture these histological features within the modeled anatomy, we created a local coordinate system in each mesh element that is aligned with the principal or mean direction of anisotropy, such as the experimentally characterized relationships between fiber angle and transmural position within the ventricular myocardium \cite{Streeter73}.
The mechanical responses of all structural components are defined by hyperelastic energy functionals, and the contractile mechanics of the myocardium are modeled by an active strain approach \cite{Ambrosi12}.
Different strain-energy functionals are used for different structures to reflect their specific material properties, and different activation waveforms are specified for the atria and the ventricles.
Supplementary Information Section \textit{\titleref{sec:klotz}} briefly discusses validation of the passive response of our left ventricle model using the methodology of Klotz et al.~\cite{Klotz06}; see Fig.~\ref{fig:klotzCurve}.

The pericardium constrains the motion of the ventricular wall, and accounting for these constraints is critical to achieve proper contraction and ventricular wall thickening during systole \cite{Pfaller19}.
As in earlier work \cite{Pfaller19}, the effect of the pericardium on myocardial movement is modeled by a parallel spring and dashpot boundary condition applied in the normal direction along the deformed epicardium.

At the length scale of the heart, blood behaves like a Newtonian fluid \cite{Peskin96}, and the dynamics of blood are well approximated by the incompressible Navier-Stokes equations.
Afterload provided by the systemic and pulmonary circulations are described using three-element Windkessel models \cite{Sturgiopulos99} applied at locations where the ascending aorta and the left and right pulmonary artery branches intersect the boundary of the computational domain.
Venous return is modeled by pressure-driven flow sources located in each atrium \cite{Griffith05}.

Calibration was primarily accomplished, as described in Methods Section \textit{\titleref{sec:bloodAndCirculation}}, by tuning the Windkessel model parameters to the outflow generated by contraction of the left ventricle via a process that mimics the baroreceptor reflex, which is a physiological control mechanism that adjusts vascular tone to maintain physiological blood pressure \cite{Hall16}.
We also adjusted the timing and magnitude of the atrial and ventricular contractions.
We emphasize, however, that many of these measures vary widely across the adult population \cite{Cattermole17}, and obtaining these statistics from both patients and healthy subjects often relies on indirect estimation (e.g., determining left ventricular volume as the volume of an ellipsoid with long axis and short axis measurements obtained via echocardiography).
Because of the high variability of these performance statistics across the adult population, it is important to ensure that our model captures intrinsic features of the cardiac cycle that cannot be summarized by simple statistics, such as the isovolumetric phases and the shapes of the blood flow rate waveforms passing through the aortic and mitral valves.

\subsection*{Cardiac Fluid Dynamics}

\begin{figure}[t!]
\begin{center}
\includegraphics[width=0.9\textwidth]{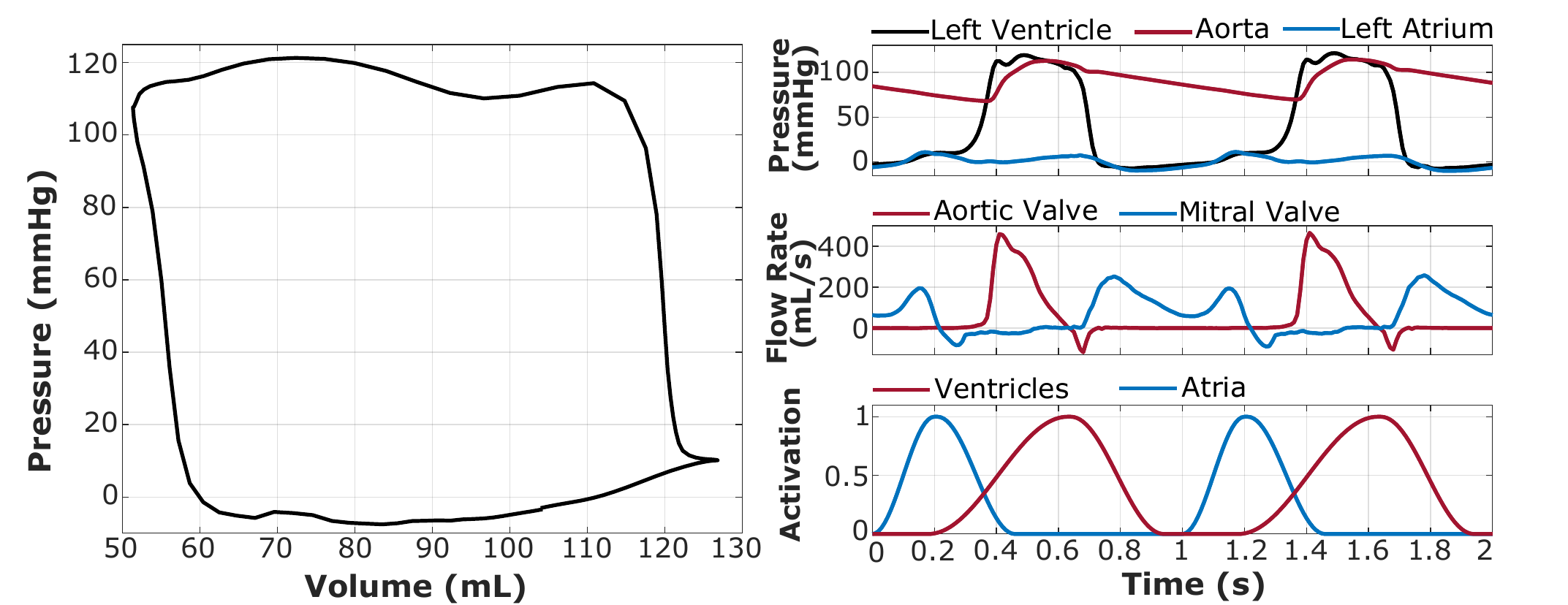}
\end{center}
\captionsetup{width=0.9\textwidth}
\caption{The left ventricular pressure-volume loop shown in the left panel captures characteristics of the cardiac cycle, including the isovolumetric phases and the stroke volume.
The right panels show pressure, flow rate, and activation waveforms for two successive cardiac cycles.
The pressure-volume loop corresponds to the second cycle shown, which is cycle five from our simulation results.}
\label{fig:pvloop}
\end{figure}

\begin{figure}[t!]
\begin{center}
\includegraphics[width=0.9\textwidth]{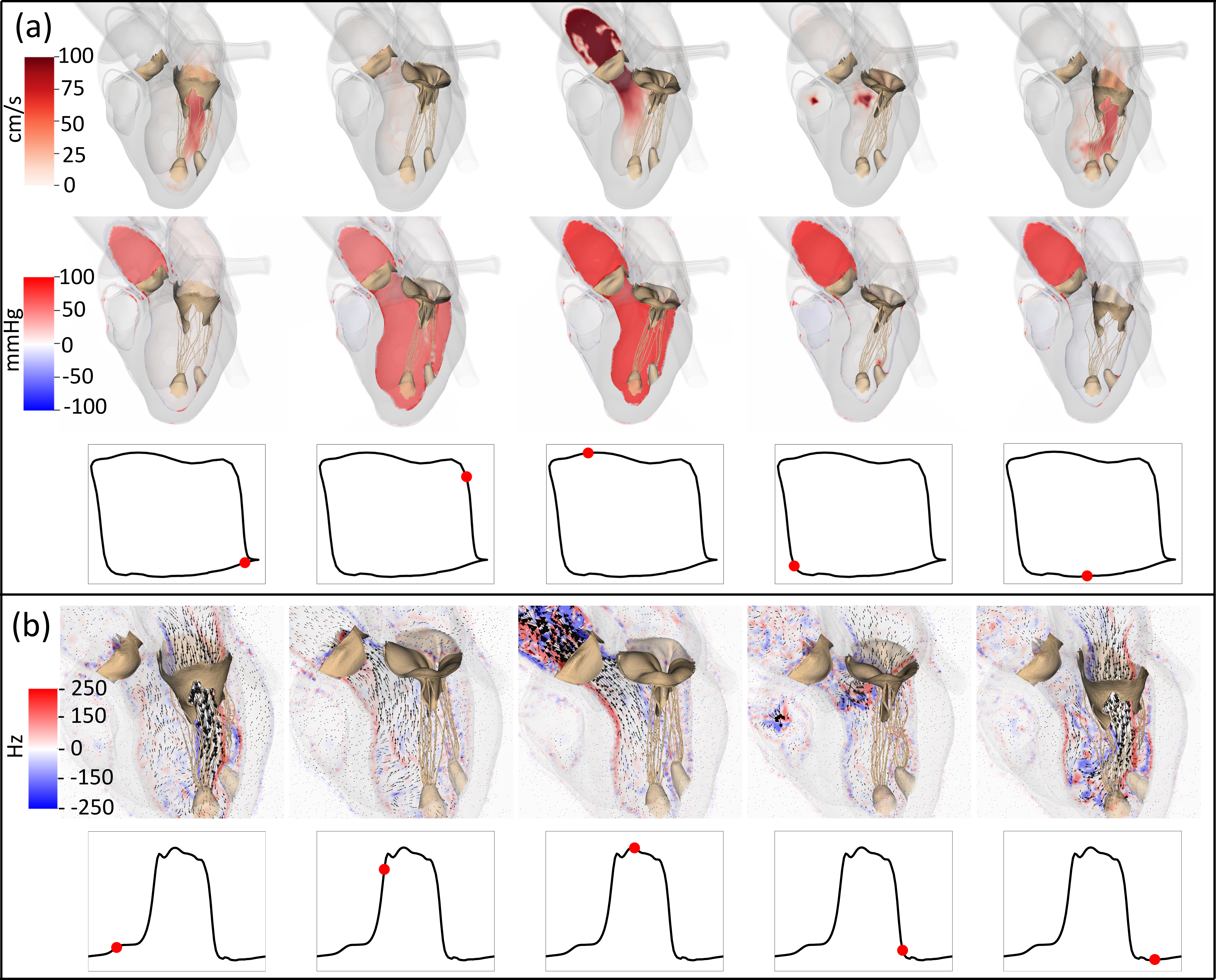}
\end{center}
\captionsetup{width=0.9\textwidth}
\caption{Cardiac fluid dynamics.
(a)~Top and middle panels show renderings of blood velocity magnitude and pressure, respectively, along a plane bisecting the aortic and mitral valves with semi-translucent renderings of the chambers at five timepoints in the cardiac cycle.
The bottom panels show the pressure-volume loop along with a red marker indicating the point in the cardiac cycle that is being visualized in the top and middle rows.
%See also Supplemental Movies 1 and 2.
(b)~Top panel shows renderings of blood velocity vectors along a plane bisecting the aortic and mitral valves with semi-translucent renderings of the chambers as well as the component of vorticity normal to the bisecting plane at five timepoints in the cardiac cycle.
The left ventricular pressure waveform is included beneath, and the red marker corresponds to the point in the cardiac cycle that is being visualized in the top row.
%See also Supplemental Movie 3.
}
\label{fig:cardiac_fluid_dynamics}
\end{figure}

\begin{table}[t!]
\centering
\begin{tabular}{|| c | c | c | c | c | c || }
\hline
   & \textbf{EDV (mL)} & \textbf{ESV (mL)} & \textbf{SV (mL)} & \textbf{EF} & \textbf{CO (L$\boldsymbol\cdot$min$^\textbf{-1}$)} \\
\hline
   \textbf{Model} & 127.58 & 52.0 & 75.58 & 0.59 & 4.5 \\
\hline
  \textbf{Reference} \cite{Hall16} & 120 & 50 & 70 & 0.58 & 4.2 \\
\hline
\end{tabular}
\captionsetup{width=0.9\textwidth}
\caption{Variables extracted from the pressure-volume loop data.
ESV = end systolic volume, EDV = end diastolic volume, SV = stroke volume, EF = ejection fraction, and CO = cardiac output.
CO is directly computed from the stroke volume and the heart rate (60 BPM).
Clinically, CO is commonly determined from oxygen saturation measurements during cardiac catheterization or estimated using Doppler echocardiography \cite{Cattermole17}.}
\label{tab:pvloop}
\end{table}

Contraction in the heart model is driven using time-periodic atrial and ventricular activation waveforms at a heart rate of 60 beats per minute (BPM).
The model reaches an approximate periodic steady state after five cycles.
The results presented here focus on left ventricular dynamics because of the large body of available clinical and experimental data.
The left ventricular pressure-volume relation, shown in Fig.~\ref{fig:pvloop}, characterizes left-ventricular performance.
The left panel shows the mean pressure sampled within the left ventricle plotted against the left ventricular volume to generate a pressure-volume loop.
The right panels of Fig.~\ref{fig:pvloop} depict pressure waveforms measured from the left atrium, left ventricle, and aorta as well as flow rate waveforms measured through the aortic and mitral valves.
Supplementary Information Section \textit{\titleref{sec:leftVentricularVolumeDynamics}} details the left ventricular volume data, shown in Fig.~\ref{fig:volume}, that are used to generate the pressure-volume relation.
Pressure-volume loop data are used to calculate end-systolic volume, end-diastolic volume, stroke volume, ejection fraction, and cardiac output, which are summarized in Table \ref{tab:pvloop}.
Reference values for a healthy adult are included for comparison.

Fig.~\ref{fig:cardiac_fluid_dynamics}(a) % and Supplemental Movies~1 and 2 show
shows the blood velocity magnitude and pressure on a plane through the left atrium, left ventricle, and part of the ascending aorta that approximately bisects the aortic and mitral valves.
These quantities are shown along with translucent renderings of the myocardium and fully opaque renderings of the aortic valve and the mitral valve apparatus.
Each of the five columns corresponds to a timepoint in the cardiac cycle, including isovolumetric contraction (second column) and relaxation (fourth column).
The red markers in the pressure-volume curves in the bottom row correspond to the same time points as the pressure and volume data in the top and middle rows for reference.
Fig.~\ref{fig:cardiac_fluid_dynamics}(b) % and Supplemental Movie~3 depict
depicts the vorticity and velocity vector field on a slice through the left heart with a time series plot of left-ventricular pressure included beneath for reference.
Periods of isovolumetric constraction and relaxation are clearly seen in the second and fourth columns, respectively.

\subsection*{Cardiac Valve Dynamics}

\begin{figure}[t!]
\begin{center}
\includegraphics[width=0.9\textwidth]{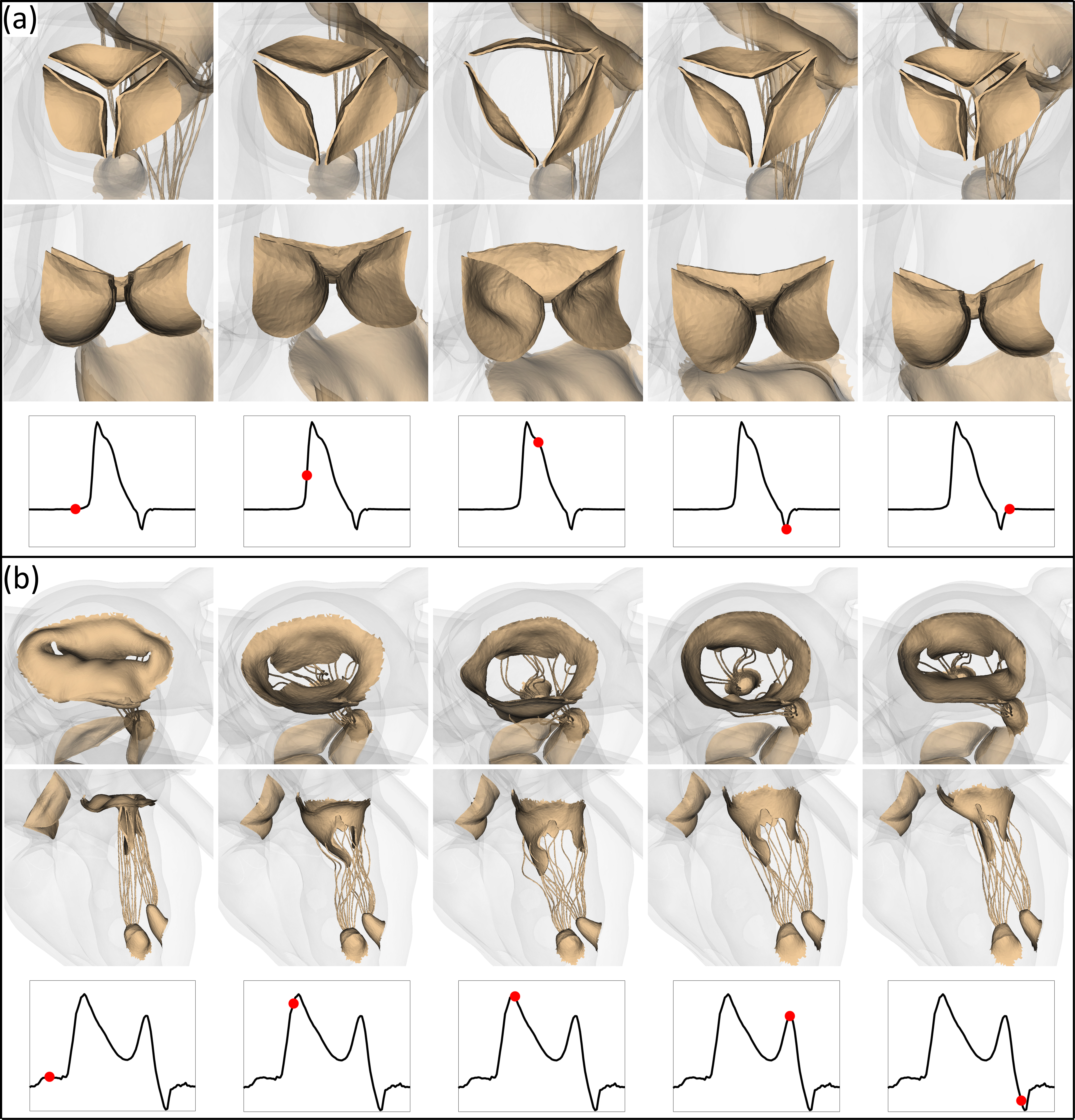}
\end{center}
\captionsetup{width=0.9\textwidth}
\caption{Cardiac valve dynamics. (a)~The top panels show the aortic valve deformations from two different views for five time points in the cardiac cycle.
The bottom panels depict the aortic flow rate waveform along with a red marker corresponding to the same timepoint visualized in the cardiac cycle.
%See also Supplemental Movie 4.
(b)~The top panels show the mitral valve deformations from two different views for five time points throughout a cardiac cycle.
The bottom set of panels depict the mitral flow rate waveform along with a red marker corresponding to the same timepoint visualized in the cardiac cycle.
%See also Supplemental Movie 5.
}
\label{fig:merged_valves}
\end{figure}

Fully three-dimensional and biomechanically detailed descriptions of the cardiac valves are key characteristics of the model.
Fig.~\ref{fig:merged_valves} % and Supplemental Movies 4 and 5 provide
provides visualizations of the deformations of the aortic and mitral valves along with flow rate waveforms measured within the valve annuli.
Both valves are shown from top and side views to highlight the opening and closing dynamics of the leaflets.
The side view of the mitral valve includes the chordae and papillary muscles to highlight their role in maintaining closure of the valve during peak ventricular systole.

\subsection*{Comparisons to Clinical and Experimental Data}

\begin{figure}[t!]
        \centering
        \includegraphics[width=0.9\textwidth]{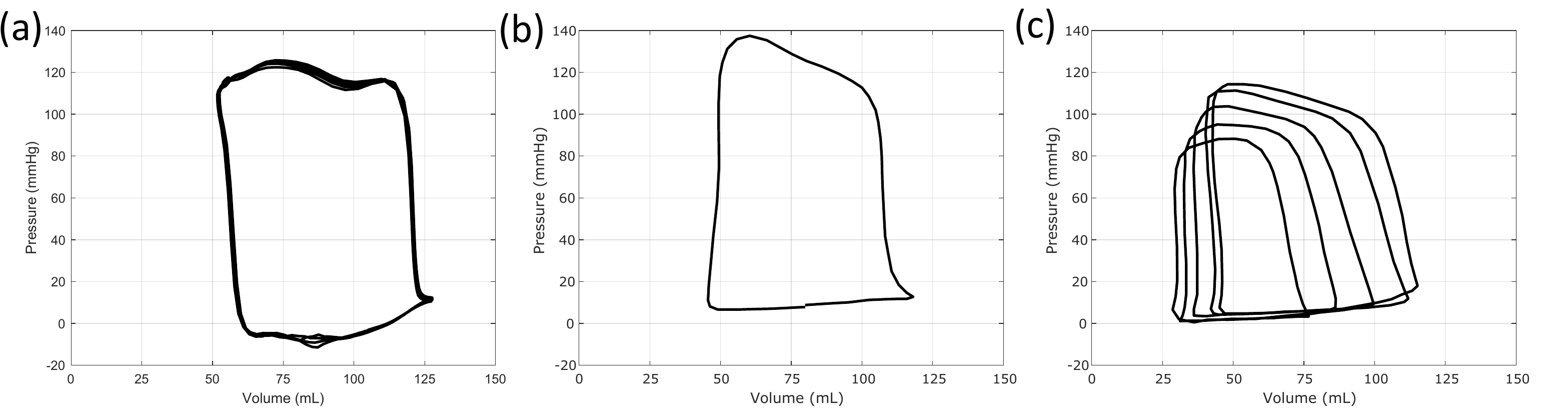}
        \captionsetup{width=0.9\textwidth}
        \caption{Successive pressure-volume curves from the fifth to eighth cycles of the model (a) compared to clinical pressure volume relations from a single cycle \cite{Patterson21} (b) and a Frank-Starling mechanism study via pulmonary vein occlusion \cite{Bastos20} (c).}
        \label{fig:pvComp}
\end{figure}

Fig.~\ref{fig:pvComp} compares the pressure-volume loops generated over successive cycles of the model to clinical pressure-volume relations \cite{Bastos20, Patterson21}.
The mitral closing transients are apparent in the cusps in the bottom right corners of the clinical pressure-volume relations, though they are not as distinct as in the model.
The closing transients for the aortic valve are not as clearly defined.
The clinical pressure-volume relations also demonstrate that the transitions in and out of the isovolumetric phases are not sharp, and the isovolumetric phases are not strictly volume preserving, which is in clear contrast to the strict vertical phases seen in textbook pressure-volume loops \cite{Hall16}.

\begin{figure}[t!]
        \centering
        \includegraphics[width=0.8\textwidth]{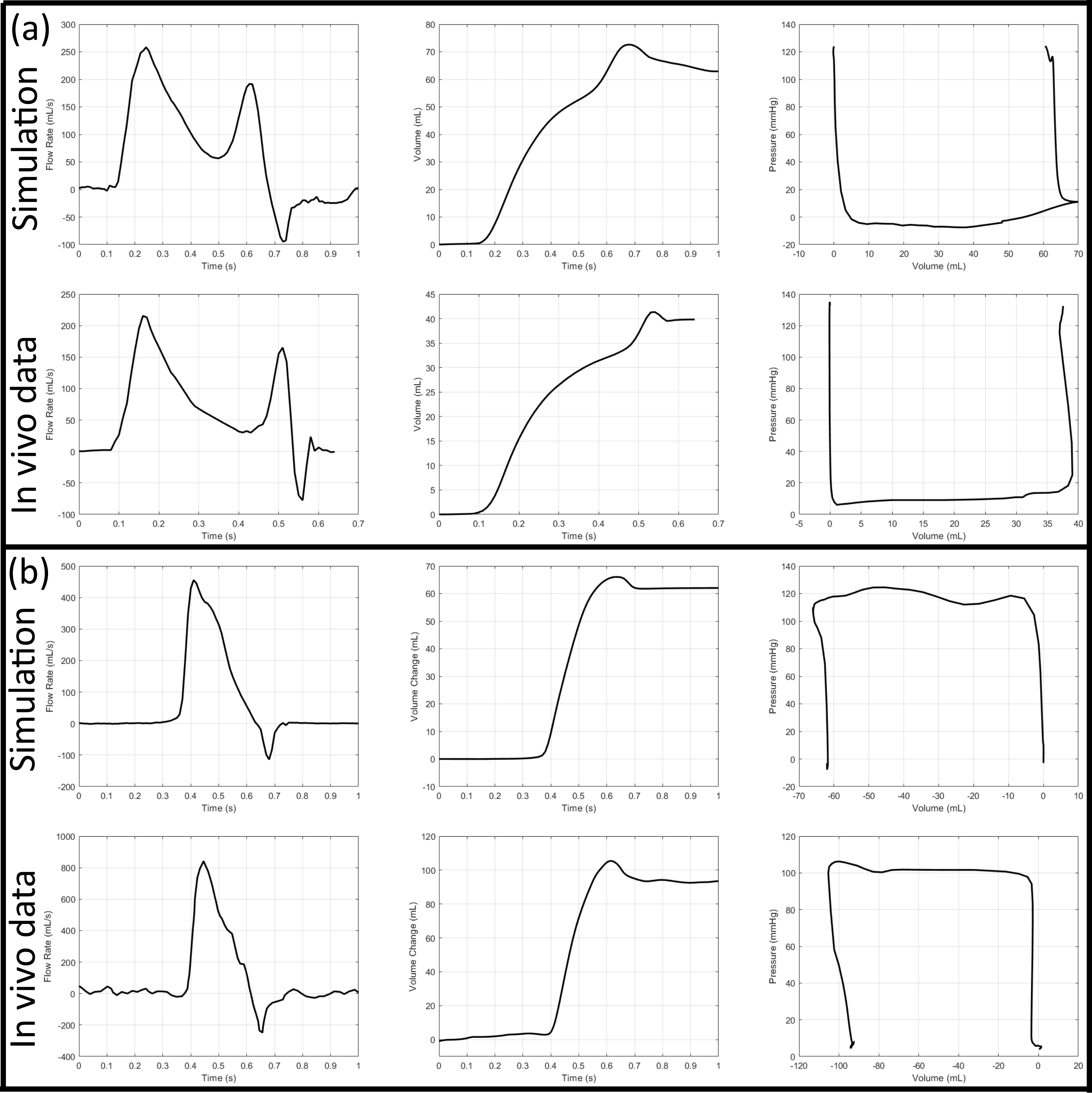}
        \captionsetup{width=0.9\textwidth}
        \caption{(a) Comparison of in vivo canine mitral valve flow rate data \cite{Yellin90} to the model.
The second column shows the volume that has passed through the mitral valve to the left ventricle during the cardiac cycle calculated by integrating the flow rate waveform.
The third column shows the mitral valve contribution to the left ventricular pressure-volume relation.
(b) Comparison of in vivo human aortic valve flow rate data \cite{Murgo80} to the model.
The second column shows the volume that has passed through the aortic valve into the aorta during the cardiac cycle calculated by integrating the flow rate waveform.
The third column shows the aortic valve contribution to the left ventricular pressure-volume relation.}
        \label{fig:mvavpv}
\end{figure}

Fig.~\ref{fig:mvavpv}(a) compares digitized in vivo canine mitral flow rate data \cite{Yellin90} to data generated by the model.
In the first column, it is apparent that the mitral flow rate waveform shape generated by the model is qualitatively similar to the in vivo data, including the mitral valve closing transient indicated by the negative flow rate.
For the simulated and experimental flow rate waveforms, we compute the volume passing through the mitral valve over one cycle by integrating the flow rate data.
The in vivo and model data look very similar in that there is a large increase in the volume output followed by a small loss in volume during the closing transient, more pronounced and prolonged in the model, and an eventual leveling of the volume.
The right column plots the left-ventricular pressure against the volume passing through the mitral valve to show the contribution of the mitral valve flux to the total pressure-volume relationship.
The in vivo pressure data come from the same source as the flow rate data, and the flow rate and pressure data were measured simultaneously \cite{Yellin90}.
The simulation and experimental data show that the isovolumetric contraction periods are not truly isovolumetric because of the mitral valve closing transient and the method used to determine volume, which does not account for the dynamic fluid volume captured between the mitral valve ring and the closed mitral valve leaflets.
These results are consistent with the clinical pressure-volume loops detailed in Fig.~\ref{fig:pvComp}.

Fig.~\ref{fig:mvavpv}(b) compares digitized in vivo human aortic flow rate data \cite{Murgo80} to data generated by the model.
In the first column it is apparent that the aortic flow rate waveform shape generated by the model is qualitatively similar to the clinical data, including the aortic valve closing transient indicated by the negative flow rate.
For both cases, we compute the volume passing through the aortic valve over one cycle by integrating the flow rate waveform.
The clinical and simulation data look very similar in that there is a large increase in the volume output followed by a small loss in volume during the closing transient and a leveling of the volume, as seen in the second column.
The right column plots the left-ventricular pressure against the volume passing through the aortic valve to show the contribution of the aortic valve flux to the total pressure-volume relationship, which shows clear volume gain during `isovolumetric' relaxation for both the model and in vivo data.
The in vivo pressure data come from the same source as the flow rate data, and the flow rate and pressure data were measured simultaneously \cite{Murgo80}.

\section*{Discussion}

The pressure-volume relationship identifies important quantitative and qualitative features of the cardiac cycle.
These include the isovolumetric periods during contraction and relaxation, when pressure is respectively increasing and decreasing in the left ventricle, but neither the aortic valve nor the mitral valve are permitting flow.
We obtain pressure-volume relationships that do not perfectly mirror pressure-volume curves seen in medical textbooks \cite{Hall16} or simulations using ideal (diode-like) valve models \cite{Watanabe04}, but that nonetheless are similar to clinical measurements appearing in the literature \cite{Bastos20, Patterson21}.
One clear feature in the pressure-volume relation generated by the model that would not be captured by a model with ideal valves is the cusp in the bottom-right of the loop, which corresponds to the mitral valve closing transient.
In vivo recordings show similar volume losses during `isovolumetric' contraction \cite{Bastos20, Patterson21}; see Figs.~\ref{fig:pvComp} and \ref{fig:mvavpv}(a).
These volume losses, however, are accompanied by a larger gain in left ventricular pressure compared to the model, highlighted in Fig.~\ref{fig:mvavpv}(a), resulting in a notable lack of a sharp cusp in the in vivo examples.
In contrast to the sharp cusp in the pressure-volume relationship generated by the mitral valve closing transient, the volume change resulting from the aortic valve transient occurs with a larger change in left ventricular pressure.
This is in good agreement with clinical aortic flow data obtained from healthy subjects \cite{Murgo80}; see Fig.~\ref{fig:mvavpv}(b).

The aortic valve closure transient seen in column 4 of Fig.~\ref{fig:merged_valves}(a) yields a small (4.01 mL) regurgitant flow volume and is in excellent qualitative agreement with human clinical data \cite{Murgo80}, as detailed in Fig.~\ref{fig:mvavpv}(b).
The mitral flow rate waveform can be quantitatively assessed using the measured deceleration time and the ratio of the peak magnitudes of the so-called E and A flow rate waveforms, which for our model are respectively 250 ms and 1.35.
These values fall within the expected ranges for healthy adults \cite{Mottram05}.
We remark, however, that clinical measurements typically use blood velocity waveforms obtained via Doppler echocardiography.
Obtaining comparable measurements from the model would require simulating the acquisition protocols used clinically.
The mitral valve flow rate waveform also captures many complex qualitative features of in vivo canine studies of mitral valve flow rates \cite{Yellin90}.
This includes a mitral valve closing transient during early ventricular contraction before the valve closes completely, highlighted in the fifth column of Fig.~\ref{fig:merged_valves}(b).
During this period in our model, a small (5.27 mL) regurgitant flow volume is lost from the left ventricle into the left atrium.
This behavior has been clearly seen in canine experimental studies, and Fig.~\ref{fig:mvavpv}(a) provides a direct comparison between our model and such experimental data.

Overall, because the dynamics of our heart model all result from mechanistic interactions among the structures of the heart and the blood, the level of agreement with existing clinical and experimental data is excellent.
Indeed, a strength of the model is that it captures the subtleties of in vivo pressure-flow relations, such as the closing transients and resultant volume changes, rather than emulating ideal pressure-volume curves that are achieved by non-compliant diode-like descriptions of the valves.
Further, a major advantage of our model is its ability to permit direct and simultaneous access to metrics such as flows through the valve annuli and localized pressures.
Such data are critical for, e.g., simulation studies of treatments for structural heart disorders such as mitral valve regurgitation.
The discussion of translational applications for this model is centered on valvular disorders because, unlike previous whole-heart models, this model has fully three-dimensional descriptions of the valves with material properties that reflect human valvular tissue.
Indeed, valvular heart diseases represent a substantial and rapidly growing burden on public health throughout the world \cite{SCoffey21}.
These include rheumatic heart disease, which primarily affects young people between the ages of 5 and 15 and causes permanent damage to one or more heart valves, along with aortic stenosis and mitral regurgitation, which are mainly diseases of the elderly.
This model is also relevant for mechanistic studies of cardiac dysfunction, such as reduced ventricular compliance from cardiomyopathy and discordant or non-extant contraction following acute myocardial infarction, and congenital heart defects, such as those within the spectrum of single ventricle physiology.
In these cases, our framework could lead to insights into experimental treatments and surgeries that can be used to improve clinical care and corresponding outcomes.

\section*{Methods}

\subsection*{Anatomical Model}

The anatomy of the heart chambers and the nearby great vessels were reconstructed from cardiac CT images (voxel size 0.32~mm$\times$0.32~mm$\times$0.4~mm) of a healthy adult male provided by Siemens Healthineers.
All data were fully deidentified by Siemens, and the study team has no way to determine the identity of the subject.
Chamber reconstruction used methods previously detailed by Segars et al.~\cite{Segars08, Segars10}.
Because the cardiac valves were obscured in the CT images, we generated idealized anatomical models of the valves with dimensions that are consistent with prior experimental studies of human heart valves.
The aortic valve leaflet geometry \cite{Hasan17} reflects the sinus height \cite{Reul90}, valve diameter \cite{Swanson74}, lunulae coaptation height \cite{Swanson74}, and leaflet thickness \cite{Clark74} of human aortic valves.
The pulmonary valve leaflets are a replica of the aortic valve leaflets that were scaled to fit within the model pulmonary artery.
The mitral valve leaflet surface geometry was built using a set of parametric equations derived from mitral valve imaging data \cite{Khalighi18}, which was further modified to match the thickness \cite{Lim05} and length \cite{Prot09} of human mitral valves.
The tricuspid valve leaflet geometry was based on valve dissection studies \cite{Lama16, Skwarek04} and has three identical leaflets that were adjusted for length to ensure closure during systole.
The papillary muscles were placed in locations identified in the CT images and connected to the atrioventricular valves by chordae tendineae.
Marginal chordae were uniformly distributed along the mitral and tricuspid valve leaflet edges, and strut and basal chordae were added to the mitral valve leaflets to prevent prolapse \cite{Khalighi19}.
These valve models were subsequently merged with the CT-derived chamber anatomy, yielding the model illustrated in Fig.~\ref{fig:model-construction}(a).

\begin{figure}[h!]
        \centering
        \includegraphics[width=0.9\textwidth]{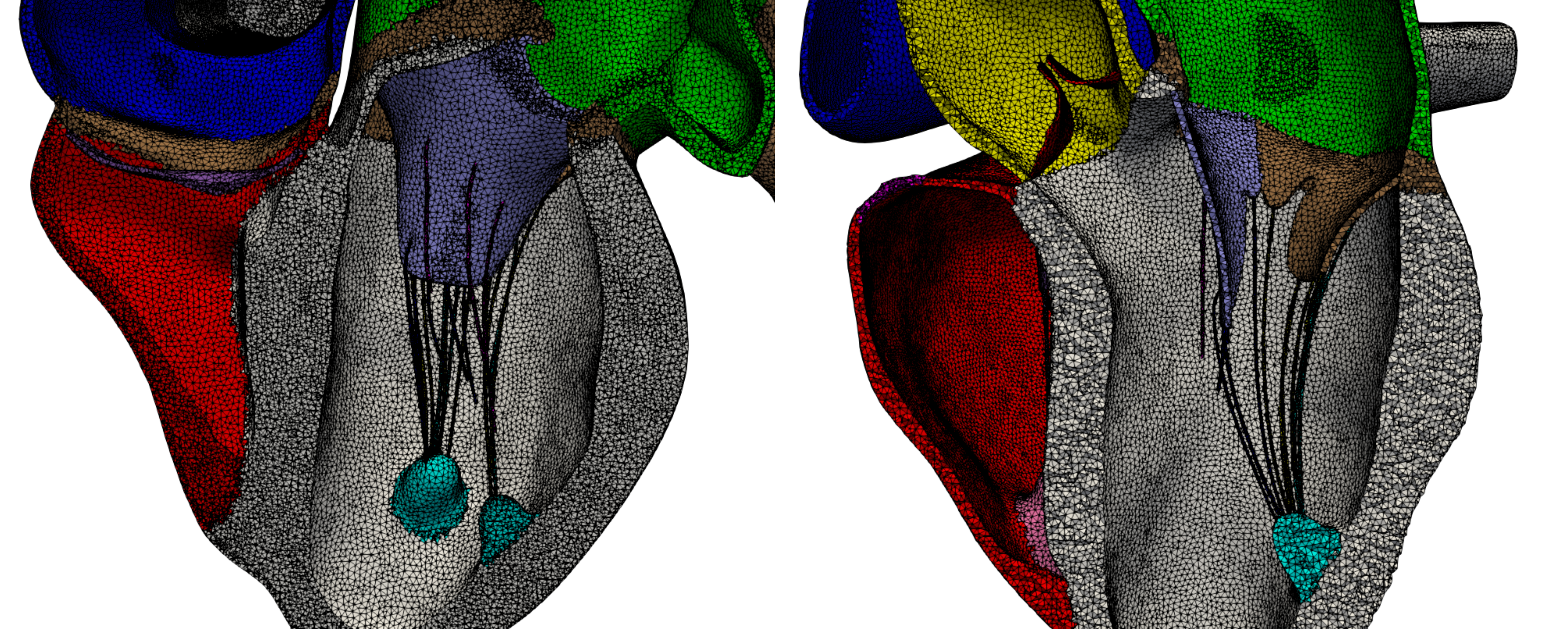}
        \captionsetup{width=0.9\textwidth}
        \caption{A visualization of the heart mesh with a focus on the left side and mitral valve apparatus.
        The mesh contains approximately 2.4 million tetrahedral elements with an average diameter of 1.17 mm.
        The colors denote different subdomains within the mesh.}
        \label{fig:meshViz}
\end{figure}

We used TetWild \cite{Hu20} to construct a monolithic, conforming tetrahedral mesh that includes the myocardium, valves, cardiac skeleton, and great vessels.
The mesh used by our model contains approximately 2.4M elements with an average diameter of 1.17 mm.
The mesh is partitioned into subdomains, which are groups of elements that share the same constitutive model and material properties.
Subdomains defined in the model include: the left and right atria; the left and right ventricles; the aortic and pulmonary valves; the mitral and tricuspid valves, including the valve cusps, chordae tendenae, and papillary muscles; the cardiac skeleton; and short segments of the great vessels, including the ascending aorta, pulmonary artery, superior and inferior vena cavae, and pulmonary veins.
Because these structures are all described within a single conforming mesh, no additional mechanical coupling conditions need to be imposed along interfaces between subdomains.
The anatomy captured in the model fits within a bounding box with dimensions 20 cm$\times$11.6 cm$\times$17.4 cm, with the longest dimension corresponding to the distance between the branch termini of the pulmonary artery.
Fig.~\ref{fig:meshViz} shows two perspectives of the heart mesh and selected subdomains, with a focus on the left ventricle and mitral valve apparatus.

\subsection*{Cardiac Biomechanics Models}
\label{sec:cardiacBiomechanics}

\subsubsection*{Fiber architecture}

To describe the local material coordinate directions that determine the tissue anisotropy, we created an orthonormal reference frame $\{\fiber{e}{f},\fiber{e}{s},\fiber{e}{n}\}$ in each element of the structural mesh.
Briefly, in the myocardium, $\fiber{e}{f}$ is the principal myofiber orientation, $\fiber{e}{s}$ points in the transmural direction (from the endocardium to the epicardium), and $\fiber{e}{n} = \fiber{e}{f} \times \fiber{e}{s}$.
The material axes in the valve leaflets capture the collagen fiber architecture identified in human heart valves \cite{Bigi82, Driessen05}.
The $\fiber{e}{f}$ direction field describes the mean collagen fiber orientation and runs from commissure to commissure in each valve, and $\fiber{e}{s}$ runs radially from the valve ring, where the leaflet intersects the myocardium, to the free edge.
To account for fiber angle dispersion within the valve leaflets, our valve biomechanics models use two distinguished collagen fiber directions, $\textbf{e}_{\text{f}^\pm} = \cos(\theta_\text{f}) \fiber{e}{f} \pm \sin(\theta_\text{f}) \fiber{e}{s}$.
These directions are rotations of $\fiber{e}{f}$ about $\fiber{e}{n} = \fiber{e}{f} \times \fiber{e}{s}$ by the angle $\pm\theta_\text{f}$, which is a material parameter fit to experimental data.
In the chordae tendineae, $\fiber{e}{f}$ is aligned with the long axis of each chord.
Fig.~\ref{fig:model-construction}(c--d) visualize the myocardial fiber directions, and Fig.~\ref{fig:model-construction}(e) shows the valve leaflets' mean collagen fiber direction with the fibers for the chordae and papillary muscles.

We use a harmonic interpolation technique that has been widely adopted in model-based approaches to describe cardiac fiber architecture \cite{Wong14}.
Specifically, on each subdomain of the structure's reference configuration $\Omegas_0$, we construct functions $\phi(\X)$ that satisfy $\nabla^2\phi(\X) = 0$, with $\X \in \Omegas_0$, and we use each resulting function to determine a local coordinate direction via $\grad\phi(\X)/\|\grad\phi(\X)\|$.
The orientation of each direction field is controlled through boundary conditions for $\phi(\X)$ that are imposed along the subdomain boundaries.
For instance, to model a group of fibers that originates on one part of the subdomain boundary and terminates along another part of the boundary, we respectively impose the Dirichlet boundary conditions $\phi(\X) = 0$ along the origin and $\phi(\X) = 1$ along the terminus.
To prevent fibers from passing through a part of the subdomain boundary, we impose homogeneous Neumann boundary conditions, $\p\phi(\X)/\p\N = 0$, in which $\N = \N(\X)$ is the unit normal to the subdomain boundary.

The ventricular material axes follow the rule-based method of Rossi et al.~\cite{Rossi14}, which reflects experimentally characterized relationships between fiber angle and transmural position within the ventricular myocardium \cite{Arts01, Nielsen91, Streeter73}.
In this approach, the sheet axis $\fiber{e}{s}$ is generated first by the harmonic interpolation procedure with boundary conditions $\phi(\X) = 0$ on the endocardium and $\phi(\X) = 1$ on the epicardium, which produces a transmurally oriented direction field.
A reference direction field that runs from the heart's apex to the mitral and tricuspid annuli through the ventricular myocardium is generated in each ventricle as $\fiber{e}{n$_0$} = \fiber{c}{} - (\fiber{c}{} \cdot \fiber{s}{0})$, in which $\fiber{c}{}$ is the vector pointing from the apex to the center of the chamber's atrioventricular valve ring.
An initial circumferential field is then created by setting $\fiber{e}{f$_0$} = \fiber{e}{s} \times \fiber{e}{n$_0$}$.
Next, $\fiber{e}{f$_0$}$ is rotated about the $\fiber{e}{s}$ axis to capture transmural fiber rotation according to rules \cite{Holzapfel09,Rossi14} based on histology studies of the ventricular myocardium \cite{Arts01, Nielsen91, Streeter73} to generate the myofiber orientation $\fiber{e}{f}$.
Finally, we set $\fiber{e}{n} = \fiber{e}{f}\times\fiber{e}{s}$.

The myofiber architecture of the atria is substantially more complex than the ventricular fiber structure, but prior studies have identified subregions within the atria with distinct principal myofiber orientations that are amenable to rule-based fiber models \cite{Ho12,Ferrer15,Krueger11}.
Herein we use a version of the rule-based approach detailed by Rossi et al.~\cite{Rossi22}.
As with the ventricles, a transmural direction field $\fiber{e}{s}$ is generated by setting the boundary conditions $\phi(\X) = 0$ on the endocardial surface and $\phi(\X) = 1$ on the epicardial surface.
The atria are then partitioned into anatomical subregions.
Within each subregion, the principal myofiber orientation is known, and subdomain boundary conditions are applied to generate the required myofiber field $\fiber{e}{f}$.
Finally, we set $\fiber{e}{n} = \fiber{e}{f}\times\fiber{e}{s}$, as in the ventricles.

The method for generating the collagen fiber network in the valves is adopted from Hasan et al.~\cite{Hasan17}.
The transmural direction field $\fiber{e}{n}$ is generated by setting $\phi(\X) = 0$ on the leaflet surface facing the ventricle and $\phi(\X) = 1$ on the leaflet surface facing the great vessels and the atria for the semilunar valves and atrioventricular valves, respectively.
A radial direction field, $\fiber{e}{s}$, is generated in each leaflet by setting $\phi(\X) = 0$ on the edge where the leaflet intersects with the myocardium and $\phi(\X) = 1$ along the leaflet's free edge.
The circumferential direction field corresponding to the mean collagen fiber axis is then determined as $\fiber{e}{f} = \fiber{e}{s}\times\fiber{e}{n}$.
The chordae tendineae fiber axes $\fiber{e}{f}$ are captured by setting $\phi(\X) = 0$ on the surfaces where the chordae meet the papillary muscles and $\phi(\X) = 1$ on the surfaces where the chordae meet the valve leaflets.

\subsubsection*{Material characterization}

The biomechanical responses of all major structures of the heart are described using the framework of large deformation elasticity.
Briefly, $\Omegas_0$ is a Lagrangian reference coordinate system attached to the initial configuration of the heart, and $\Omegas_t$ is the current configuration at time $t$.
The deformation mapping $\boldsymbol\chi:(\Omegas_0,t)\mapsto\Omegas_t$ relates reference and current coordinates, so that $\boldsymbol\chi(\X,t) \in \Omegas_t$ is the current position of $\X \in \Omegas_0$ at time $t$.
The mechanical responses of all structural components are defined by hyperelastic energy functionals $\W$ of the deformation gradient tensor $\FF(\X,t) = \p\boldsymbol\chi(\X,t)/\p\X$ \cite{Holzapfel09}.
We describe myocardial contractile mechanics using an active strain approach \cite{Ambrosi12}, which assumes that the Helmoltz free energy $\W$ can be expressed using $\FF$ and an internal variable $\FFA$ that represents the active component of the deformation, yielding $\W = \W(\FF, \FFA)$.
The active strain model links $\FF$ and $\FFA$ through an intermediate virtual configuration, so that $\FF = \FFE \FFA$, and it assumes that the energy can be defined in the intermediate configuration, such that $\W(\FF, \FFA) = \W(\FFE) = \W(\FF \FFA^{-1})$ \cite{Rossi14}.

Different strain-energy functionals are used for different structures to reflect their specific material properties.
Following the principle of material objectivity, the hyperelastic models are formulated using the right Cauchy-Green strain tensor, $\CC = \FF\tran \FF$, in terms of $I_1 = \text{tr}(\CC)$, $I_{4i} = \textbf{e}_i\tran\CC \textbf{e}_i$, $I_{4i}^\star = \max(I_{4i},1)$, and $I_{8ij} = \textbf{e}_i\tran \CC \textbf{e}_j$, in which $i$ and $j$ index the material coordinate axes.

The ventricles and papillary muscles use the orthotropic Holzapfel-Ogden model \cite{Holzapfel09},
\begin{equation} \label{eq:W_HO}
        \W = \frac{a}{2b}\exp(b(I_1-3)) +
        \sum_{i=\text{f,s}} \frac{a_{i}}{2b_{i}}(\exp(b_{i}(\kappa_{i} I_1 + (1-3\kappa_{i})I_{4i} - 1)^2)-1) +
        \frac{a_\text{fs}}{2b_\text{fs}}(\exp(b_\text{fs}I_{8\text{fs}}^2)-1).
\end{equation}
We use material parameters from G\"{u}ltekin et al.~\cite{Gultekin16} that are based on triaxial shear tests on cuboid specimens of human left ventricles.
The atria use the model of Augustin et al.~\cite{Augustin19},
\begin{equation} \label{eq:W_Augustin}
        \W = \frac{a}{2b}(\exp(b(I_1-3))-1) +
        \frac{a_\text{f}}{2b_\text{f}}(\exp(b_\text{f}(\kappa I_1 + (1-3\kappa)I_{4\text{f}} - 1)^2)-1).
\end{equation}
We use material parameters from Augustin et al.~\cite{Augustin19} that were calibrated using biaxial strain test data from anterior and posterior specimens of human left atria.
The parameters $\kappa_i$ in Eq.~\eqref{eq:W_HO} and $\kappa$ in Eq.~\eqref{eq:W_Augustin} characterize myofiber angle dispersion \cite{Holzapfel09, Augustin19}.

\begin{figure}[h!]
        \centering
        \includegraphics[width=0.9\textwidth]{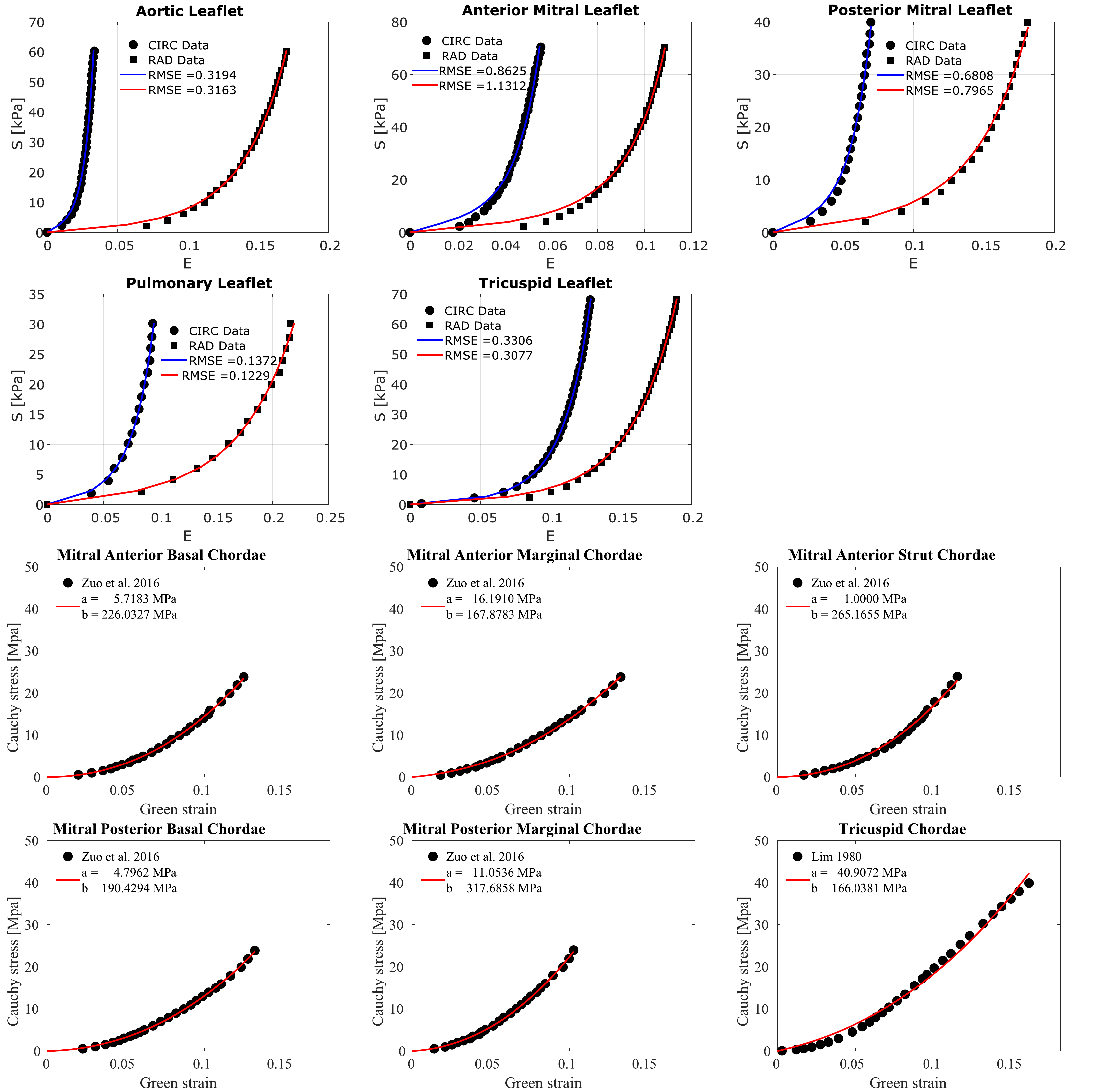}
        \captionsetup{width=0.9\textwidth}
        \caption[this is here to avoid errors in multibib]{Parameters for the valve material models were derived from the results of biaxial stress-strain tests along the radial and circumferential directions executed by Pham et al.~\cite{Pham17}, and fit to a modified Holzapfel-Gasser-Ogden model proposed by Murdock et al.~\cite{Murdock18}, Eq.~\eqref{eq:W_HGO}.
        Data from mitral \cite{Zuo16} and tricuspid \cite{Lim80} chordae stress-strain tests were fit to a generic nonlinear spring material model, Eq.~\eqref{eq:W_chordae}.
        Digitized data are shown as the black circles and squares for circumferential and radial fiber directions, respectively, for the valves.
        The fits for the radial and circumferential data are shown as the red and blue lines, respectively, along with the corresponding root mean squared errors (RMSE) for the valve fits.
        The digitized data for the chordae are shown as black circles, and the fits are shown as the red curves with the resultant parameters listed.}
        \label{fig:valvesChordaeFits}
\end{figure}

The valve leaflets use a version of the Holzapfel-Gasser-Ogden model \cite{Holzapfel00} by Murdock et al.~\cite{Murdock18},
\begin{equation}\label{eq:W_HGO}
        \W = a\{\exp[b(I_1-3)]-1\} +
        \frac{a_\text{f}}{2b_\text{f}}\sum_{k\in\{+,-\}}\{\exp[b_{\text{f}}(I_{4\text{f}^k}^\star-1)^2]-1\}.
\end{equation}
We fit the parameters for all valves using biaxial stress-strain data generated by Pham et al.~\cite{Pham17}, as described below.
Notice that the collagen fiber stresses in our leaflet models engage in tension but not in compression, which corresponds to the concept that collagen fibers collapse under compression and do not substantially contribute to the stress response of the material \cite{Gasser06}.
The chordae tendineae use a nonlinear spring model,
\begin{equation} \label{eq:W_chordae}
    \W = \dfrac{a}{2}\left(I_1-3\right) + \dfrac{a_\text{f}}{3} \left( I_{4\text{f}}^\star - 1\right)^3.
\end{equation}
We determined mitral chordae parameters for the posterior and anterior leaflet chordae from uniaxial stress-strain tests of human mitral chordae \cite{Zuo16}, as described below.
We also determined tricuspid chordae parameters from uniaxial stress-strain tests of human tricuspid chordae \cite{Lim80}; all tricuspid chordae use the same parameters because of limited availability of human tissue data.
Briefly, to determine material parameters, we extracted stress-strain curves using WebPlotDigitizer \cite{Rohatgi22} and fit the constitutive model using \texttt{lsqcurvefit} in MATLAB (MathWorks, Natick, MA) with a tolerance of 1e-12.
Fig.~\ref{fig:valvesChordaeFits} shows our model fits.

\begin{table}
\hspace*{\fill}
\begin{sideways}
\begin{minipage}{\textheight}
\centering
\small
\begin{tabular}{||c | c | c | c | c |c |c |c |c |c |c |c||}
\hline
\textbf{Tissue} &$\boldsymbol{a}$ \textbf{[kPa]}& $\boldsymbol{b}$ & $\boldsymbol{a}_\textbf{f}$ \textbf{[kPa]} & $\boldsymbol{b}_\textbf{f}$ & $\boldsymbol \kappa_\textbf{f}$ & $\boldsymbol\theta_\textbf{f}$ \textbf{[rad]}& $\boldsymbol{a}_\textbf{s}$ \textbf{[kPa]} & $\boldsymbol{b}_\textbf{s}$ & $\boldsymbol\kappa_\textbf{s}$ & $\boldsymbol{a}_\textbf{fs}$ \textbf{[kPa]} & $\boldsymbol{b}_\textbf{fs}$ \\
\hline
Ventricles \cite{Gultekin16}& 0.4 & 6.55 & 3.05 & 29.05 & 0.08 & -- & 1.25 & 36.65  & 0.09 & 0.15 & 6.28 \\
\hline
Papillary Muscles \cite{Gultekin16}& 0.4 & 6.55 & 3.05 & 29.05 & 0.08 & -- & 1.25 & 36.65  & 0.09 & 0.15 & 6.28 \\
\hline
Atria \cite{Augustin19}& 2.92 & 5.6 & 11.84 & 17.95 & 0.17 & -- & -- & -- & -- & --& -- \\
\hline
Aortic Valve \cite{Pham17}& 0.172 & 25.24 & 10.2 & 384.6 & -- & 0.1055 & -- & -- & -- & --& -- \\
\hline
Anterior Mitral Leaflet \cite{Pham17}& 0.398 & 14.91 & 11.1  & 104.9 & -- & 0.7746 & -- & -- & -- & -- & --\\
\hline
Posterior Mitral Leaflet \cite{Pham17}& 0.24 & 15.63 & 2.462 & 84.1 & -- & 0.0 & -- & -- & -- & -- & -- \\
\hline
Pulmonary Valve \cite{Pham17}& 0.2135 & 11.14 & 0.9634 & 49.66 & -- & 0.0001 & -- & -- & -- & -- & --\\
\hline
Tricuspid Valve \cite{Pham17}& 0.1502 & 15.23 & 0.3157 & 40.73 & -- & 0.0058 & -- & -- & -- & -- & --\\
\hline
Anterior Mitral Basal Chordae \cite{Zuo16}& 5.718e3 & -- & 2.26e5 & -- & -- & -- & --& -- & -- & -- & --\\
\hline
Anterior Mitral Marginal Chordae \cite{Zuo16}& 1.619e4 & -- & 1.679e5 & -- & -- & -- & --& -- & -- & -- & --\\
\hline
Anterior Mitral Strut Chordae \cite{Zuo16}& 1.0e3 & -- & 2.652e5 & -- & -- & -- & --& -- & -- & -- & --\\
\hline
Posterior Mitral Basal Chordae \cite{Zuo16}& 4.796e3 & -- & 1.904e5 & -- & -- & -- & --& -- & -- & -- & --\\
\hline
Posterior Mitral Marginal Chordae \cite{Zuo16}& 1.105e4 & -- & 3.177e5 & -- & -- & -- & --& -- & -- & -- & -- \\
\hline
Tricuspid Valve Chordae \cite{Lim80}& 4.091e4 & -- & 1.66e5 & -- & -- & -- & -- & -- & -- & -- & -- \\
\hline
Aorta & 1.2e2 & -- & -- & -- & -- & -- & -- & -- & -- & -- & -- \\
\hline
Vena Cavae & 1.43e2 & -- & -- & -- & -- & -- & -- & -- & -- & -- & -- \\
\hline
Pulmonary Artery & 1.0e2 & -- & -- & -- & -- & -- & -- & -- & -- & -- & -- \\
\hline
Pulmonary Veins & 1.43e2 & -- & -- & -- & -- & -- & -- & -- & -- & -- & -- \\
\hline
\end{tabular}
\captionsetup{width=0.9\textwidth}
\caption{The numerical bulk modulus $\boldsymbol{\beta_s = 4.0}$\textbf{e4 kPa} throughout the entire structure.
Citations are included for the origin of the data that were used for the parameter fitting or the parameters themselves.
As stated in the main text, the material parameters for the great vessels are used to prevent gross deformations and are not based on human tissue studies.}
\label{tab:allParameters}
\end{minipage}
\end{sideways}
\hspace{\fill}
\end{table}

To avoid severe mitral valve regurgitation, we prestrained some chordae along their major fiber axes.
This was necessary because the model construction process produced some loose chordae, which led to valve prolapse during ventricular systole.
We used a prestraining approach that is equivalent to the active strain formulation for the myocardium \cite{Ambrosi12}, but with fiber stretch $\gamma_\text{f}$ constant in time.
The fiber stretch terms for the eighteen mitral valve chordae range in values from 0.0 to 0.25, with the majority set to 0.15.

The vena cavae, pulmonary veins, ascending aorta, and pulmonary artery are modeled as neo-Hookean materials,
\begin{equation} \label{eq:W_neoHookean}
        \W = \frac{a}{2}(I_1 - 3).
\end{equation}
Model parameters were chosen to allow for realistic vessel compliance while avoiding excessive deformation across the cardiac cycle.

Table \ref{tab:allParameters} lists all constitutive model parameters.

\subsubsection*{Active contraction}
\label{sec:activeContraction}

In the active strain formulation, $\FFA$ defines the time-dependent change in the reference configuration resulting from muscle contraction \cite{Ambrosi12}.
We use
\begin{equation}
    \FFA = \II
         + \gf{f}\, \dyad{e}{f}{e}{f}
         + \gf{s}\, \dyad{e}{s}{e}{s}
         + \gf{n}\, \dyad{e}{n}{e}{n},
\end{equation}
in which $\gamma_i$ defines the deformation scaling along each material axis $i \in \{\text{f,s,n}\}$.
The contraction of the myocardium is assumed to be volume-preserving, so $\det(\FFA) = 1$, and, for simplicity, transversely isotropic along the fiber axis $\fiber{e}{f}$.
We prescribe $\gf{f}(t)$ and thereby obtain $\gf{n}(t) = \gf{s}(t) = \left(1+\gf{f}(t)\right)^{-1/2} - 1$.

\begin{table}[t!]
\begin{center}
\begin{tabular}{|| c | c | c | c | c | c ||}
\hline
\textbf{Chamber} & \textbf{$t_\textbf{delay}$ (s)} & \textbf{$t_\textbf{peak}$ (s)} & \textbf{$t_\textbf{plateau}$ (s)} & \textbf{$t_\textbf{drop}$ (s)} &  $\boldsymbol\gamma_\textbf{f,max}$ \\
\hline
Atria & 0.0 & 0.2 & 0.01 & 0.25 &  0.08 \\
\hline
Ventricles & 0.18 \cite{Houthuizen11} & 0.45 & 0.01 & 0.3 & 0.3 \\
\hline
\end{tabular}
\captionsetup{width=0.9\textwidth}
\caption{Parameters for the active strain approach in the myocardium. The contraction period is 1 second.}
\label{tab:contractionParameters}
\end{center}
\end{table}

The contraction timings for the atria and ventricles are based on studies of conduction propagation through the atria, atrioventricular node, and ventricles \cite{Houthuizen11}, as well as pressure profiles measured within the chambers \cite{Murgo80}.
We defined the activation waveform by the function
\begin{equation}
    g(t) =
    \begin{cases}
    \frac{1}{2}-\frac{1}{2}\cos\Big(\frac{\pi}{t_\text{peak}}t\Big) & \text{if } t - t_\text{delay} < t_\text{delay}, \\
    1 & \text{if t}_\text{peak} \leq t - t_\text{delay}\leq t_\text{plateau}+t_\text{peak}, \\
    \frac{1}{2}+\frac{1}{2}\cos\left(\frac{\pi}{t_\text{drop}}(t-t_\text{peak}-t_\text{plateau})\right) & \text{if } t_\text{plateau}+t_\text{peak} < t -  t_\text{delay} < t_\text{plateau}+t_\text{peak}+t_\text{drop}, \\
    \end{cases}
    \label{eq:contractionShape}
\end{equation}
in which $t_\text{delay}$ is the time from the beginning of the cycle to the start of contraction, $t_\text{peak}$ is the time to peak contraction after the onset of contraction,  $t_\text{plateau}$ is the time spent at peak contraction, and $t_\text{drop}$ is the time from the end of peak contraction to no contraction.
Together with the magnitude of peak contraction, $\gf{f,max}$, we have $\gf{f}(t) = \gf{f,max}\, g(t)$.
The active strain parameters for the atria and the ventricles are stated in Table \ref{tab:contractionParameters}, and the respective waveforms are visualized in the bottom right panel of Fig.~\ref{fig:pvloop}.
The majority of the parameters were acquired through manual calibration except for the ventricular $t_\text{delay}$, which was chosen to correspond to the time it takes for the activation signal to propagate from the sinoatrial node to the ventricles \cite{Houthuizen11}.

\subsection*{Pericardium}

Following the approach of Pfaller et al.~\cite{Pfaller19}, the force from the pericardium, $\F(\X,t)$, imposed on the epicardial surface, is determined by a system of distributed damped springs via
\begin{equation}
        \F(\X,t) = \n(\X,t) \otimes \n(\X,t) \left[\kappa\,(\X - \boldsymbol\chi(\X,t)) - \eta\, \U(\X,t)\right],
        \label{eq:pericardium}
\end{equation}
in which $\n(\X,t)$ is the surface unit normal to the epicardium in the current configuration, $\U(\X,t)$ is the velocity in the current configuration of material point $\X$, $\kappa$ is a tethering constant, and $\eta$ is a damping constant.
Fig.~\ref{fig:model-construction}(b) shows a schematic of the pericardium model superimposed with the full heart geometry.
In our simulations, the parameters for the pericardium model are $\kappa = 1.0$ kPa/mm and $\eta = 5.0\text{e-}2$ kPa$\cdot$s/mm.
These parameters were chosen to limit gross epicardial oscillations, especially during ventricular relaxation.

\subsection*{Blood and Circulation}
\label{sec:bloodAndCirculation}

At the length scale of the heart, blood behaves like a Newtonian fluid \cite{Peskin96}, and the dynamics of blood are well approximated by the incompressible Navier-Stokes equations.
We choose uniform mass density $\rho = 1.0 \text{ g} \cdot \text{cm}^{-3}$ and uniform dynamic viscosity $\mu = 4 \text{ mPa}\cdot\text{s}$ \cite{Brindise18}.

The afterload provided by the systemic and pulmonary circulations are described using three-element Windkessel models applied at locations where the ascending aorta (Ao) and the left and right pulmonary artery branches (LPA and RPA) intersect the boundary of the computational domain \cite{Griffith12, Sturgiopulos99}.
The state variable for each of these models is the afterload pressure, downstream of the great vessel, that satisfies the equation
\begin{align}
C_i\dfrac{{\mathrm d}p_{\text{wk},i}}{{\mathrm d}t} + R_{\text{p},i}\, p_{\text{wk},i} = Q_{\text{outflow},i}, \quad i \in \{\text{Ao, LPA, RPA}\}.
\end{align}
$Q_{\text{outflow},i}$ is the volume of blood per unit time leaving vessel $i$ at the boundary of the computational domain.
The Windkessel parameters for the great vessels were chosen to be consistent with clinical measurements \cite{Murgo80, Hall16, Naeije13}.
Model calibration was performed by querying the outflow rate waveforms and adjusting the peripheral resistance and compliance terms to maintain physiologic pressures at the boundary \cite{Murgo80} following a procedure described below.
The models of the peripheral circulations used in this study are not closed, and flows and pressures from the left and right sides of the heart are uncoupled. Instead, venous return is modeled by a pressure-driven flow source located in each atrium (LA and RA) \cite{Griffith05}.
The flow sources are determined by
\begin{align}
L_j\dfrac{{\mathrm d}Q_j}{{\mathrm d}t} + R_j\,Q_j = p_{\text{source},j} - p_{\text{atrium},j}, \quad j \in \{\text{LA, RA}\},
\end{align}
in which $p_{\text{source},j}$ is the pressure upstream of the respective atrium, which is treated as constant, and $p_{\text{atrium},j}$ is the pressure sampled within the chamber.
The ratio of the parameters $L_j$ and $R_j$ governs the timescale of the flow response to the difference between the source and atrium pressures.
Fig.~\ref{fig:model-construction}(b) shows the Windkessel circulation models and the flow sources in relation to the full heart geometry.

\begin{table}[t!]
\begin{center}
\begin{tabular}{|| c | c | c | c ||}
\hline
\textbf{Vessel} & $\boldsymbol{R_\textbf{C}}$ \textbf{(mmHg$\boldsymbol\cdot$s/mL)} & $\boldsymbol{R_\textbf{P}}$ \textbf{(mmHg$\boldsymbol\cdot$s/mL)}& $\boldsymbol{C}$ \textbf{(mL/mmHg)}\\
\hline
Aorta & 0.033 \cite{Sturgiopulos99} & 1.46 & 0.7 \\
\hline
Pulmonary Arteries & 0.0219 & 0.08 & 5.56 \\
\hline
\end{tabular}
\captionsetup{width=0.9\textwidth}
\caption{The systemic circulation Windkessel parameters are tuned to the flow rate output at the edge of the ascending aorta.
The pulmonary circulation Windkessel parameters are based on in vivo pressure measurements.}
\label{tab:wkParameters}
\end{center}
\end{table}

\begin{figure}[h!]
        \centering
        \includegraphics[width=0.9\textwidth]{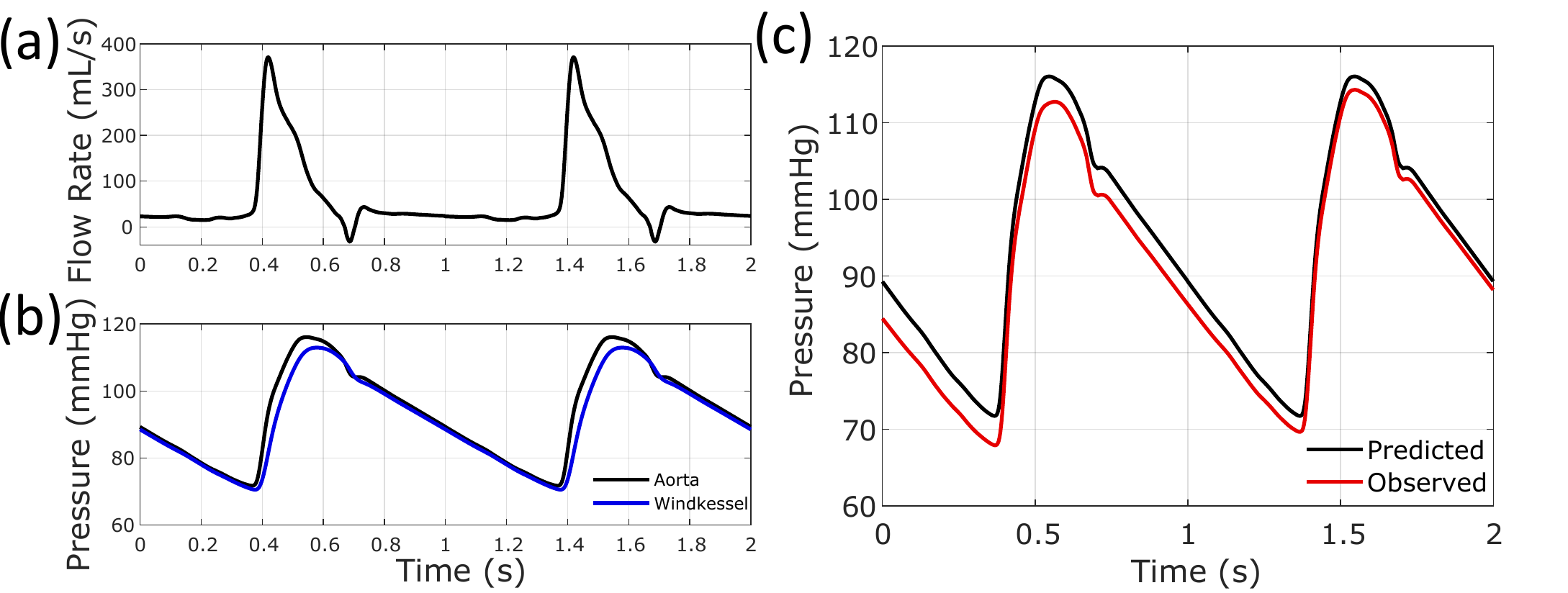}
        \captionsetup{width=0.9\textwidth}
        \caption{Workflow for tuning the Windkessel parameters for the aorta.
        (a) The tuning of the Windkessel parameters starts with extracting the flow rate waveforms from the top of the aorta.
        (b) The flow rate is then used to choose the compliance and peripheral resistance that lead to appropriate systolic and diastolic pressures in the aorta.
        (c) The model is then run with these parameters and the observed pressure values are validated against the predicted pressure waves. Over successive cycles, the observed pressure converges to the pressure wave predicted by the specified Windkessel model parameters.}
        \label{fig:WKFit}
\end{figure}

Parameters for the aorta and the pulmonary arteries are provided in Table \ref{tab:wkParameters}.
The characteristic resistance value for the aorta was taken from Sturgiopoulos et al.~\cite{Sturgiopulos99}.
The peripheral resistance and compliance were calibrated to fit specific flows generated from contraction of the left ventricle.
The first step of this calibration procedure was to calculate the flow rate waveform at the intersection of the aorta with the edge of the computational domain; see Fig.~\ref{fig:WKFit}(a).
The parameters were then adjusted to generate physiologic systolic and diastolic pressure values \cite{Murgo80}.
Fig.~\ref{fig:WKFit}(b) shows the Windkessel model predictions of the pressure waveform in the aorta as well as the pressure waveform downstream from the characteristic resistance.
The model fit was tested, as illustrated in Fig.~\ref{fig:WKFit}(c), by comparing the predicted pressure from the Windkessel model and the observed pressure at the intersection of the aorta and the computational domain.
Over successive cycles, the pressure range generated by the model converges to the pressure wave predicted by the specified Windkessel model parameters.
This procedure mimics the baroreceptor reflex, which is a physiological control mechanism that adjusts vascular tone to maintain physiological blood pressure \cite{Hall16}.

Parameters for the pulmonary arteries were based on pulmonary circulation values.
To calculate nominal values for these parameters, we assumed a cardiac output of 100 mL/s.
With the heart rate defined to be 60 BPM, the stroke volume is 100 mL.
The pulmonary arteries were assumed to have a systolic pressure of 19 mmHg and diastolic pressure of 10 mmHg, with a mean pulmonary arterial pressure of 13 mmHg and a mean pulmonary venous pressure of 9 mmHg \cite{Naeije13}.
We assumed the mean pressure drop from the pulmonary arteries to the pulmonary veins is completely described by the peripheral resistance, and that the flow rate was equally split between the two pulmonary arteries.
This resulted in a nominal value for the peripheral resistance of $R_\text{p} = 0.08$ mmHg$\cdot$s/mL.
The compliance value was calculated as the fraction of the stroke volume entering the artery divided by the pulse pressure.
This resulted in a compliance value of 5.56 mL/mmHg.
The pulmonary characteristic resistance was chosen to be of comparable magnitude to but smaller than the aortic characteristic resistance.

The flow source parameters were also determined empirically.
The inertance of each source, which we have found to be most important with regards to maintaining numerical stability, is chosen to be time-step size dependent, and it is set to be as small as possible while preventing spurious changes in flow rate \cite{Griffith05}.
The resistances for both the left and right atrial sources are 0.15 mmHg$\cdot$s/mL.
For the right atrial source, the inertance is 300$\cdot\Delta t$ mmHg$\cdot$s$^2$/mL and the pressure source is 3.75 mmHg, where $\Delta t$ is the time step size.
For the left atrial source, the inertance is 240$\cdot\Delta t$ mmHg$\cdot$s$^2$/mL and the pressure source is 10 mmHg.

\subsection*{Fluid-Structure Interaction}

Our model uses an immersed approach to simulating fluid-structure interaction (FSI).
The immersed boundary (IB) method \cite{Peskin02}, originally introduced by Peskin to model the fluid dynamics of heart valves \cite{Peskin72}, is the earliest example of such a numerical method.
It treats fluid-structure systems in which an elastic structure is immersed in a viscous incompressible fluid.
The IB method describes the structure in Lagrangian form and the fluid in Eulerian form, and it uses integral equations with Dirac delta function kernels to connect the Lagrangian and Eulerian frames.
When the governing equations are discretized for computer simulation, the singular delta function is replaced by a regularized version of the delta function.
Our computations use an efficient nodal version \cite{Wells23} of a stabilized immersed finite element/difference (IFED) method \cite{Griffith17, Lee22, VadalaRoth20, Wells23}.
This scheme is a variation on the IB method that uses a finite element description of the structure, enabling structural models with complex geometries and realistic constitutive models.
The IFED method also uses a regularized version of the Dirac delta function, and the choice of delta function used in this study follows results from a recent benchmarking study \cite{Lee22}.
The remainder of this section outlines the IFED method and provides details on numerical discretization parameters used to generate simulation results.

Briefly, the IFED method predicts the coupled dynamics of the fluid-structure system within a computational domain $\Omega$ that is partitioned into non-overlapping fluid and solid subdomains, $\Omegaf_t$ and $\Omegas_t$, that are indexed by time $t$.
To enable the use of fast structured-grid solvers, we require that $\Omega = \Omegaf_t \cup \Omegas_t$ is a fixed rectangular region.
Our simulations use a computational domain of size 20 cm$\times$20 cm$\times$20 cm, which is slightly larger than the bounding box that contains the reconstructed anatomy.
The IFED formulation uses both Eulerian variables, which are described using fixed physical coordinates $\x \in \Omega$, and Lagrangian variables, which are described using material coordinates $\X$ that are chosen to be the initial coordinates of the structure at time $t=0$, so that $\X \in \Omegas_0$.
The deformation mapping $\vchi: (\Omegas_0,t) \mapsto \Omegas_t \subseteq \Omega$ connects reference and current coordinates, so that $\vchi(\X,t) \in \Omegas_t$ is the current position of material point $\X$ at time $t$.

The equations of motion for the coupled fluid-structure system are
\begin{align}
  \label{eq:nse}
  \rho \left(\dfrac{\p \u}{\p t}(\x, t) + \u(\x,t) \cdot \nabla \u(\x,t) \right)
  &= -\nabla p(\x,t) + \mu \nabla^2 \u(\x,t) + \f(\x, t), & \text{$\x \in \Omega$}, \\
  \nabla \cdot \u(\x, t) &= 0,  & \text{$\x \in \Omega$}, \\
  \label{eq:continuous-forces}
  \f(\x, t) &= \int_{\Omegas_0} \F(\X, t) \, \delta(\x - \vChi(\X, t)) \,\DX, & \text{$\x \in \Omega$}, \\
  \label{eq:continuous-velocity}
  \dfrac{\p \vChi}{\p t}(\X, t)
  &= \U(\X, t)
  = \int_\Omega \u(\x, t) \, \delta(\x - \vChi(\X, t)) \,\Dx,  & \text{$\X \in \Omegas_0$},
\end{align}
in which $\u(\x,t)$ and $\U(\X, t)$ are Eulerian and Lagrangian velocity fields, $p(\x,t)$ is the pressure, $\f(\x,t)$ and $\F(\X,t)$ are Eulerian and Lagrangian elastic force densities, $\rho$ is the mass density of the fluid-structure system, $\mu$ is the viscosity, and $\delta(\x)$ is the Dirac delta function.
The Lagrangian elastic force density is defined in terms of the first Piola-Kirchhoff structural stress tensor, $\PP(\X,t) = \dfrac{\p\W}{\p\FF}(\X,t)$, by requiring $\F(\X,t)$ to satisfy
\begin{equation}
  \label{eq:continuous-force-definition}
  \int_{\Omegas_0} \F(\X, t) \cdot \boldsymbol{\psi}(\X) \,\DX = -\int_{\Omegas_0} \PP(\X, t) : \nabla_\X \boldsymbol{\psi}(\X) \,\DX
\end{equation}
for all smooth vector-valued test functions $\boldsymbol{\psi}(\X)$.
See Boffi et al.~\cite{Boffi08} and Griffith and Luo \cite{Griffith17} for additional discussion.

As detailed in Methods Section \textit{\titleref{sec:cardiacBiomechanics}}, the biomechanical response of the heart, its valves, and the nearby great vessels are described using hyperelastic constitutive models that are formulated using elastic energy functionals $\W$ of invariants and pseudo-invariants of the right Cauchy-Green strain, $\CC = \FF\tran \FF$, in which $\FF = \p\vchi/\p\X$ is the deformation gradient tensor and $J = \det(\FF)$ is the Jacobian determinant.
Although the continuum formulation of the IFED method generates exactly incompressible deformations, for which $J \equiv 1$, this property is generally lost when the continuous equations are discretized because of both spatial and temporal discretization effects.
To mitigate these errors, we have found that it is beneficial to adopt a nearly incompressible material formulation \cite{VadalaRoth20}.
To do so, it is convenient to introduce the so-called modified Cauchy-Green strain, $\overline{\CC} = \overline{\FF}\tran \overline{\FF}$ with $\overline{\FF} = J^{-\frac13} \FF$.
Notice that $\det(\overline{\FF}) = 1$, so $\overline{\CC}$ encodes only deviatoric deformations but not dilatational motions.
We denote by $\overline{\W}$ elastic energies that use invariants of $\overline{\CC}$, i.e., in terms of the modified invariant $\overline{I}_1 = \text{tr}(\overline{\CC})$ instead of $I_1 = \text{tr}(\CC)$.
We do not use modified pseudo-invariants, because doing so can result in non-physical deformations \cite{Sansour08}.
In addition, we introduce a volumetric energy,
\begin{equation} \label{eq:W_dilatational}
  \mathcal{U}(J) = \beta_\text{s}(J\log(J) - J + 1),
\end{equation}
which penalizes changes in volume, in which $\beta_\text{s}$ is the numerical bulk modulus \cite{VadalaRoth20}.
The numerical bulk modulus $\beta_\text{s}$ is set to 4.0e4 kPa, and the same value is used throughout the entire structure.
We use $\overline{\W}$ and $\mathcal{U}$ to evaluate the first Piola-Kirchhoff elastic stress via
\begin{equation}
  \PP = \dfrac{\p \overline{\mathcal W}}{\p\FF} + \dfrac{\p {\mathcal U}}{\p\FF}.
\end{equation}
See Vadala-Roth et al.~\cite{VadalaRoth20} for further details.

In our simulations, we use an adaptive staggered-grid discretization of the incompressible Navier-Stokes equations detailed by Griffith \cite{Griffith12} and a finite element description of Lagrangian equations describing the deformation the immersed structures and the resulting force generation \cite{Griffith17}.
The interaction equations, Eqs.~\eqref{eq:continuous-forces} and \eqref{eq:continuous-velocity}, are discretized using an efficient nodal coupling scheme \cite{Wells23}, and we replace the singular delta function with a three-point B-spline kernel that was found to provide excellent accuracy and robustness compared to other commonly used choices \cite{Lee22}.
The locally refined Cartesian grid is comprised of two nested grid levels with a refinement ratio of four between levels, and it provides a fine-grid spatial resolution of 0.43 mm.
We use second-order centered differences for the Eulerian divergence, gradient, and Laplace operators along with a high-resolution upwind scheme for the convective terms \cite{Griffith12}.
We use standard $P^2$ (quadratic) tetrahedral elements to describe the structure.
The Eulerian and Lagrangian variables are coupled using an explicit midpoint method \cite{Griffith12}, and the incompressible Navier-Stokes equations are discretized in time using a semi-implicit scheme that uses the Crank-Nicolson method for the viscous terms and the second-order Adams-Bashforth scheme for the convective terms.
We use a time-step size of 2.69 $\mu$s, which was chosen to be as large as possible while avoiding volumetric instability in the structure.

\subsection*{Flow Rate, Chamber Volume, and Pressure}
\label{sec:flowMeters}

We identify valve annuli upstream of each of the four heart valves, and we construct surface meshes where blood velocities are sampled to evaluate volumetric flow rates through each valve.
The flow rate is captured every 100 time steps.
Flow volumes associated with each valve are obtained by integrating the flow rate via the trapezoidal rule.
After the volume changes are computed, the flow rate data and volume contributions are down sampled to every 0.01 s for plotting output.
The volumes that pass through the aortic valve and mitral valve are used in conjunction with the initial volume of the ventricle to compute the evolving left ventricular volume.
The initial volume of the left ventricle was determined by extracting a surface mesh of the left ventricular endocardium that was then capped by the mitral and aortic valve annular meshes.
The mesh manipulation and volume computation were done using Meshmixer (Autodesk, San Rafael, CA).
The left ventricular pressure is approximated by querying the pressure at the centroid of the left ventricular endocardial surface at every time step.
Pressure data are down sampled to every 0.01 s and smoothed using a three point moving average by the \texttt{smooth} function in MATLAB (MathWorks, Natick, MA).

\subsection*{Data Availability}

All data supporting the conclusions of this study are included in the manuscript and supporting information.

\subsection*{Code Availability}

The model was built using the IBAMR software infrastructure available on GitHub (\url{ibamr.github.io}), which is released under a permissive open-source license.
This software relies on SAMRAI \cite{Hornung02} for its finite difference framework, libMesh \cite{Kirk06} for its finite element framework, and PETSc \cite{PETSc0, PETSc1, PETSc2} for its linear solver infrastructure.

\renewcommand\refname{References}
\bibliographystyle{pnas-new}
\bibliography{heart_model_refs}

\section*{Acknowledgements}

We acknowledge research funding from National Institutes of Health Awards R01\-HL117063, U01\-HL143336, and R01\-HL157631 and National Science Foundation Awards DMS 1460368, OAC 1460334, OAC 1450327, OAC 1652541, and OAC 1931516.
MD was supported in part by National Institutes of Health Institutional Training Grant Award T32\-GM067553.
CP was supported in part by National Science Foundation Research Training Group Award DMS 1646339.
Computations were performed using facilities provided by the University of North Carolina at Chapel Hill through the Research Computing division of UNC Information Technology Services.
BEG, CSP, and CP acknowledge discussions and contributions by David M.~McQueen in the development of the immersed boundary method and its applications to predecessors of the model of cardiac fluid dynamics described in this study.

\section*{Author Contributions}

\textbf{Marshall Davey:}~Methodology; Software; Validation; Formal Analysis; Investigation; Data Curation; Writing - Original Draft; Writing - Review \& Editing; Visualization.
\textbf{Charles Puelz:}~Conceptualization; Methodology; Software; Validation; Investigation; Writing - Original Draft; Writing - Review \& Editing.
\textbf{Simone Rossi:}~Methodology; Software; Investigation; Writing - Review \& Editing; Visualization.
\textbf{Margaret Anne Smith:}~Formal Analysis; Investigation; Data Curation.
\textbf{David R.~Wells:}~Methodology; Software; Writing - Original Draft; Writing - Review \& Editing.
\textbf{Greg Sturgeon:}~Methodology; Software; Formal Analysis; Investigation; Data Curation; Writing - Review \& Editing.
\textbf{Paul Segars:}~Methodology; Software; Formal Analysis; Investigation; Data Curation; Resources; Writing - Review \& Editing.
\textbf{John P.~Vavalle:}~Validation; Writing - Review \& Editing.
\textbf{Charles S.~Peskin:}~Methodology; Validation; Writing - Review \& Editing.
\textbf{Boyce E.~Griffith:}~Conceptualization; Methodology; Software; Validation; Resources; Writing - Original Draft; Writing - Review \& Editing; Supervision; Project Administration; Funding Acquisition.
MD and CP contributed equally to this work.
BEG is the corresponding author.

\section*{Competing Interests}

The authors declare no competing interests.

\section*{Materials and Correspondence}

Correspondence and material requests should be addressed to Boyce E.~Griffith at \url{boyceg@email.unc.edu}.

\clearpage

\beginsupplement

\section*{SUPPLEMENTARY INFORMATION}

\section*{Left Ventricular End-Diastolic Pressure-Volume Relationship}
\label{sec:klotz}

\begin{figure}[h!]
        \centering
        \includegraphics[width=0.9\textwidth]{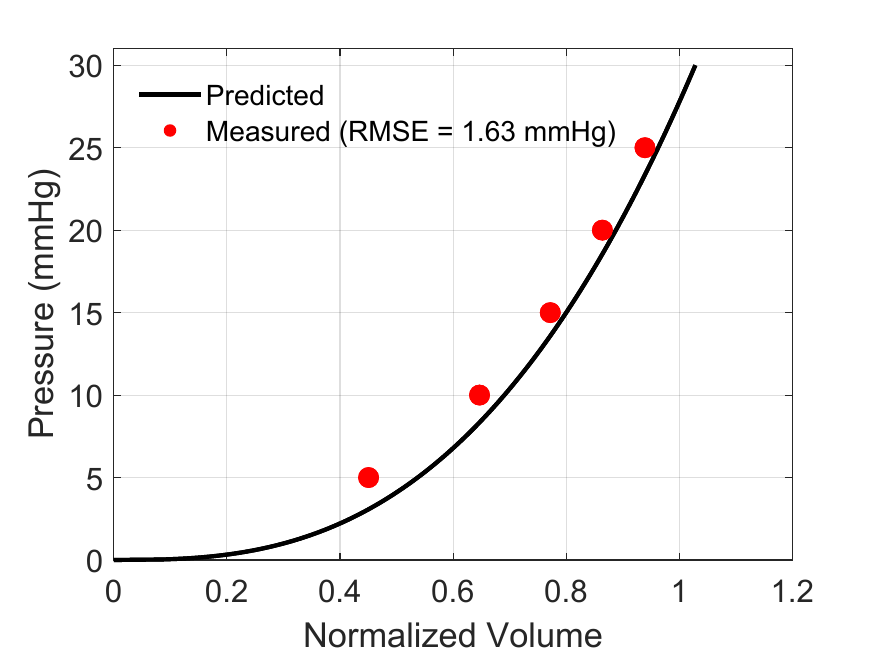}
        \captionsetup{width=0.9\textwidth}
        \caption{The Klotz curve generated from our model, which was used to verify the passive material parameters within the left ventricle \cite{Klotz06}.
        Pressures of 5, 10, 15, 20, 25, and 30 mmHg were applied to the left ventricular endocardial surface with pericardial tethering removed from the epicardial surfaces.
        The resultant left ventricular volumes were 115.8, 124.3, 129.8, 133.8, 137.0, and 139.7 mL, respectively, with an initial left ventricular volume of 96.3 mL.
        Compared to the predicted curve estimated (black curve) by Klotz et al., the root mean squared error of the observed data (red dots) is 1.63 mmHg, which is well within to the observed in vivo human data range of 2.99$\pm$1.72 mmHg \cite{Klotz06}.}
        \label{fig:klotzCurve}
\end{figure}

To provide a partial validation of the passive elastic response of the model heart, we generated the Klotz pressure-volume curve as detailed by Klotz et al.~\cite{Klotz06}.
To do so, we removed the pericardial tethering from the ventricular epicardium, applied loads of 5, 10, 15, 20, 25, and 30 mmHg to the endocardial surface of the left ventricle, and recorded the resulting chamber volume.
The Klotz relation relies on a normalized volume $\tilde{v}(p)$ defined by
\begin{equation}
        \tilde{v}(p) = \frac{v(p) - v_0}{v_{30} - v_0},
\end{equation}
in which $v(p)$ is the volume (mL) of the left ventricle at pressure $p$ (mmHg), $v_{30}$ is the volume of the left ventricle at 30 mmHg, and $v_0$ is the volume of the left ventricle at 0 mmHg.
Fig.~\ref{fig:klotzCurve} shows the Klotz curve data generated by our model.
We obtain good agreement to the reference curve, with a root mean squared error of 1.63 mmHg.
This error is comparable to the root mean squared error of 2.99$\pm$1.72 mmHg reported by Klotz et al.~for in vivo human data \cite{Klotz06}.

\section*{Left Ventricular Volume Dynamics}
\label{sec:leftVentricularVolumeDynamics}

\begin{figure}[h!]
        \centering
        \includegraphics[width=0.9\textwidth]{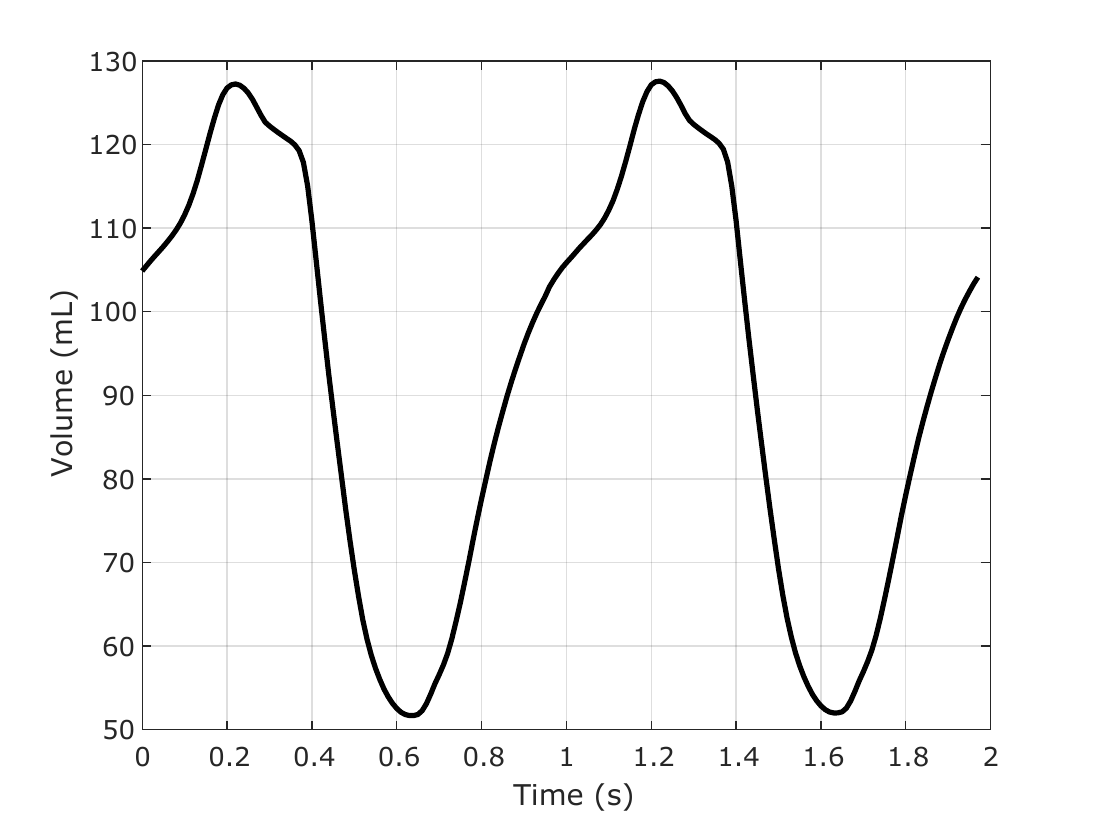}
        \captionsetup{width=0.9\textwidth}
        \caption{
        The total left ventricular volume during two successive cycles.}
        \label{fig:volume}
\end{figure}

Fig.~\ref{fig:volume} shows the time series data for the left ventricular volume that were used to generate the pressure-volume relation.
The pressure data are shown in the main body of the text, but volume waveforms are much less commonly obtained in vivo.
As mentioned in Methods Section \textit{\titleref{sec:flowMeters}}, the left ventricular volume was estimated using the fluxes through the mitral and aortic valves along with an initial volume that was determined using numerical quadrature.

\newpage

\newpage

\newpage

\newpage

\end{document}